\documentclass[onecolumn,nofootinbib,showpacs,preprintnumbers,amsmath,amssymb]{revtex4-1}
\usepackage{amsfonts}
\usepackage{amsmath}
\usepackage{amssymb}
\usepackage{amstext}
\usepackage{amsthm}
\usepackage{graphicx}
\usepackage{dcolumn}
\usepackage{bm}
\usepackage{color}
\usepackage{float}

\begin{document}

%\preprint{APS/123-QED}

\title{Binary versus non-binary information in real time series:\\ empirical results and maximum-entropy matrix models}

\author{Assaf Almog}
\affiliation{Instituut-Lorentz for Theoretical Physics, Leiden Institute of Physics, University of Leiden, Niels Bohrweg 2, 2333 CA Leiden (The Netherlands)}
% \altaffiliation[Also at ]{Physics Department, XYZ University.}%%%Lines break automatically or can be forced with \\
\author{Diego Garlaschelli}%
\affiliation{Instituut-Lorentz for Theoretical Physics, Leiden Institute of Physics, University of Leiden, Niels Bohrweg 2, 2333 CA Leiden (The Netherlands)}
%

%\affiliation{

\date{\today}% It is always \today, today,
\begin{abstract}
The dynamics of complex systems, from financial markets to the brain, can be monitored in terms of multiple time series of activity of the constituent units, such as stocks or neurons respectively.
While the main focus of time series analysis is on the magnitude of temporal increments, a significant piece of information is encoded into the binary projection (i.e. the sign) of such increments.
In this paper we provide further evidence of this by showing strong nonlinear relations between binary and non-binary properties of financial time series. 
These relations are a novel quantification of the fact that extreme price increments occur more often when most stocks move in the same direction.
We then introduce an information-theoretic approach to the analysis of the binary signature of single and multiple time series. 
Through the definition of maximum-entropy ensembles of binary matrices and their mapping to spin models in statistical physics, we quantify the information encoded into the simplest binary properties of real time series and identify the most informative property given a set of measurements. 
Our formalism is able to accurately replicate, and mathematically characterize, the observed binary/non-binary relations.
We also obtain a phase diagram allowing us to identify, based only on the instantaneous aggregate return of a set of multiple time series, a regime where the so-called `market mode' has an optimal interpretation in terms of collective (endogenous) effects, a regime where it is parsimoniously explained by pure noise, and a regime where it can be regarded as a combination of endogenous and exogenous factors.
Our approach allows us to connect spin models, simple stochastic processes, and ensembles of time series inferred from partial information. 
\end{abstract}

\maketitle

\section{Introduction}

In large systems, the observed dynamics or activity of each unit can be represented by a discrete time series providing a sequence of measurements of the state of that unit. 
One of the main challenges researchers are faced with is that of extracting meaningful information from the high-dimensional (multiple) time series characterizing all the elements of a complex system \cite {Mantegna&Stanley,Econophysics book,Bouchaud1,Bouchaud2,Mantegna(a),Schneidman,Dal,Mant,Onnela}. 
Traditionally, the main object of time series analysis is the characterization of patterns in the amplitude of the increments of the quantities of interest.
Given a signal $s_i(t)$ where $i$ denotes one of the $N$ units of the system and $t$ denotes one of the $T$ observed temporal snapshots, the generic increment or `return' $r_i(t)$ can be defined as 
\begin{equation}
r_i(t)\equiv s_i(t+1)-s_i(t)\qquad i=1, N\quad t=1, T
\label{eq:r}
\end{equation}
and generates a new time series.

While a time series of increments encapsulates all the relevant information about the amplitude of the fluctuations of the original signal, a significant part of this information is encoded in the purely `binary' projection of $r_i(t)$, i.e. its sign 
\begin{equation}
x_i(t)\equiv \textrm{sign} [r_i(t)]=
\left\{\begin{array}{rr}
+1&r_i(t)>0\\
0&r_i(t)=0\\
-1&r_i(t)<0
\end{array}\right.
\label{eq:sign}
\end{equation}
Previous analyses, mainly in the field of finance, have indeed documented various forms of statistical dependency between the sign and the absolute value of fluctuations, e.g. sign-volume correlations \cite{lillo, shlomo} and the leverage effect \cite{Black, MacroNews2, Bekaert, Leverage Effect}.
Other studies have also documented that the binary projections of various financial \cite{boguna} and neural \cite{hertz} time series exhibit nontrivial dynamical features that resemble those of the original data.
All these results suggest that binary projections indeed retain a non-trivial piece of information about the original time series, and call for a deeper analysis of the problem. 

Being binary, the sign of the increments is much more robust to noise than the increments themselves. Moreover, it is scale-invariant (i.e. independent of the chosen unit of increments) and does not depend on whether the original data have been preliminarily rescaled or log-transformed (as usually done e.g. for financial time series).
Binary time series can also be analyzed with the aid of much simpler mathematical models than required by non-binary data (several examples of such models will be provided in this paper).
Finally, as we show later on, in multiple financial time series the total binary increment of a given cross section measures the instantaneous level of synchronization (i.e. the number of stocks moving in the same direction) of the market, while the total non-binary increment does not carry this piece of information.
For all the above reasons, it is important to further investigate whether the full `weighted' or `valued' information can, in some circumstances, be somehow mapped to the binary one, thus providing a robust, highly simplified, more easily modeled, and informative representation of the system.

Motivated by the above considerations, in this paper we further study, both empirically and theoretically, the relationship between weighted time series and their binary  projections. 
We first provide robust empirical evidence of novel relationships between binary and non-binary properties of real financial time series.
To this end, we use the daily closing prices of all stocks of three markets (S$\&$P500, FTSE100 and NIKKEI225) over the period 2001-2011. 
We show that the average daily increment and average daily coupling of an empirical set of multiple time series are strongly and non-linearly related to the corresponding average increment of the binary projections of the same time series. These empirical relations quantify in a novel way the strong correlations existing between the increments of individual stocks and the overall level of synchronization among all stocks in the market.

Building on this evidence, we then introduce a formalism to analytically characterize random ensembles of single and multiple time series with desired constraints. 
Specifically, we follow Jaynes' interpretation and re-derivation of statistical physics as an inference problem from partial macroscopic information to the unobservable microscopic configuration \cite{Jaynes1, Jaynes2}. 
We define statistical ensembles of matrices that maximize Shannon's entropy \cite{Shannon48}, subject to a set of desired constraints. 
This maximum-entropy approach is widely used in many areas, from neuroscience \cite{neural} to social network analysis \cite{wasserman} (and more recently network science in general \cite{Park J and Newman}), where it is known under the name of ERG (Exponential Random Graph) formalism.
In the case of interest here, we introduce ensembles of maximum-entropy binary matrices that represent projections of single and multiple binary time series, subject to a set of desired constraints defined as simple empirical measurements. 
We discuss the main differences between our matrix ensembles and other techniques in time series analysis, including other ensembles of random matrices encountered in random matrix theory \cite{Mehta, MehtaB, Wigner, Dyson, Sengupta}.

Our approach leads to a family of analytically solved null models that allow us to quantify the amount of information encoded in the chosen constraints, i.e. the selected observed properties of the binary projections of real time series. 
Different choices of the constraints lead to different stochastic processes, a result that allows us to relate known stochastic processes to the corresponding `target' empirical properties defining the ensemble of time series spanned by the process itself.
After applying the approach to the financial time series in our analysis, we compare the informativeness of various measured properties and show that different properties are more relevant for different time series and temporal windows. 
We also identify distinct regimes in the behaviour of multiple stocks and give the most likely explanation (endogenous, exogenous, or mixed) for the observed level of coordination or `market mode', given the measured binary return at a given point in time.
Finally, and most importantly, we show that our approach is able to reproduce and mathematically characterize the observed non-linear relationships between binary and non-binary properties of real time series. 

The rest of the paper is  organized as follows. 
In sec. \ref{sec:data} we describe the data and provide empirical evidence of the relationships that motivate our work.
In sec. \ref{sec:methods} we introduce our theoretical formalism in its general form. 
In sec. \ref{sec:sts} we apply the formalism to single time series, while in sec. \ref{sec:smts} we apply it to single cross-sections (temporal snapshots) of multiple time series. 
Finally, in sec. \ref{sec:full} we consider our method in its full extent and apply it to entire spans of multiple time series, for different financial markets around the globe.  
We end with our conclusions in sec. \ref{sec:Conclusions}. 

\section{Empirical results\label{sec:data}}
\subsection{Data}
We use daily closing prices, for the 10-years period ranging from 24/10/2001 to 18/10/2011, of all stocks from the indices S$\&$P500,  FTSE100 and NIKKEI225.
For each index, we restrict our sample to the maximal group of stocks that are traded continuously throughout the selected period.
This results in 445  stocks for the S$\&$P500, 78 stocks for the  FTSE100 and 193  stocks for the NIKKEI225.

%%%%%%%%%%%%%
\begin{figure}[t]
\includegraphics[width=.4\textwidth]{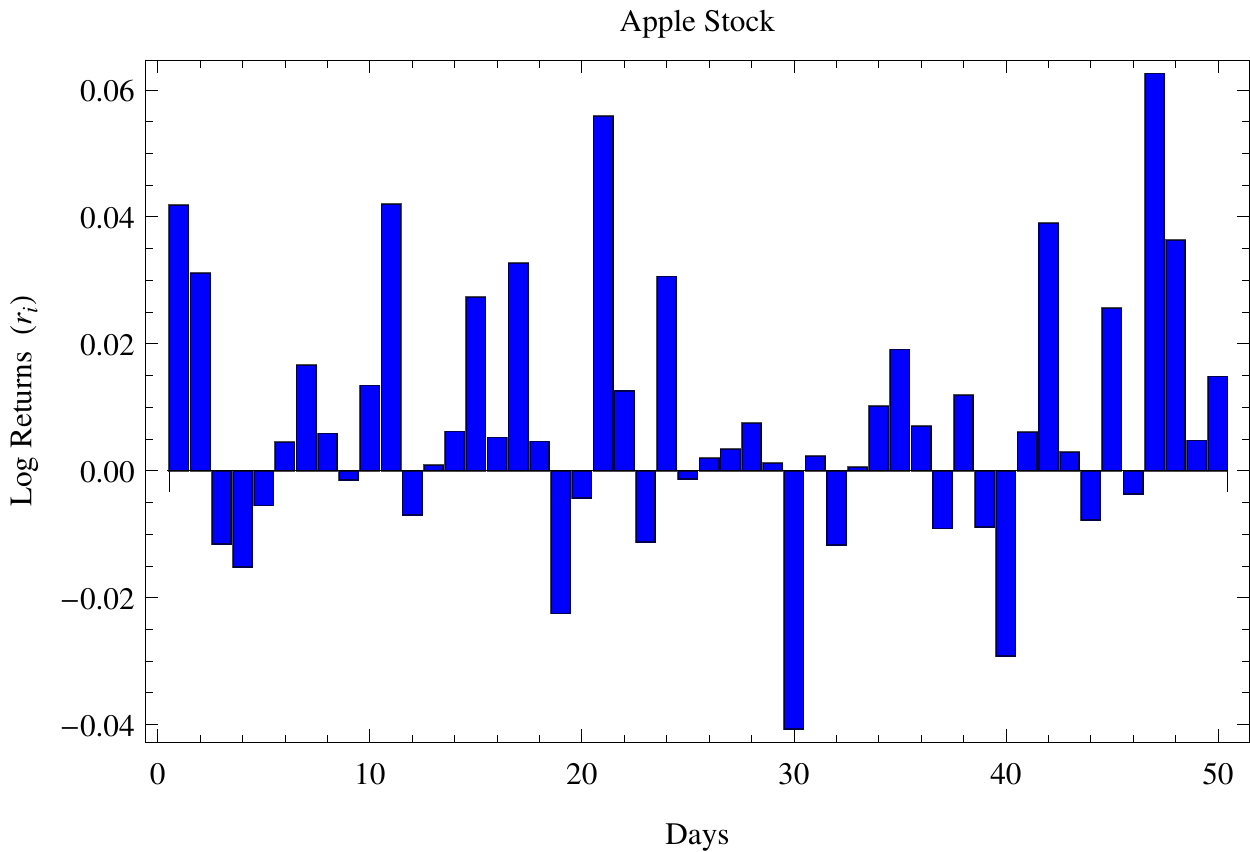}
\includegraphics[width=.4\textwidth]{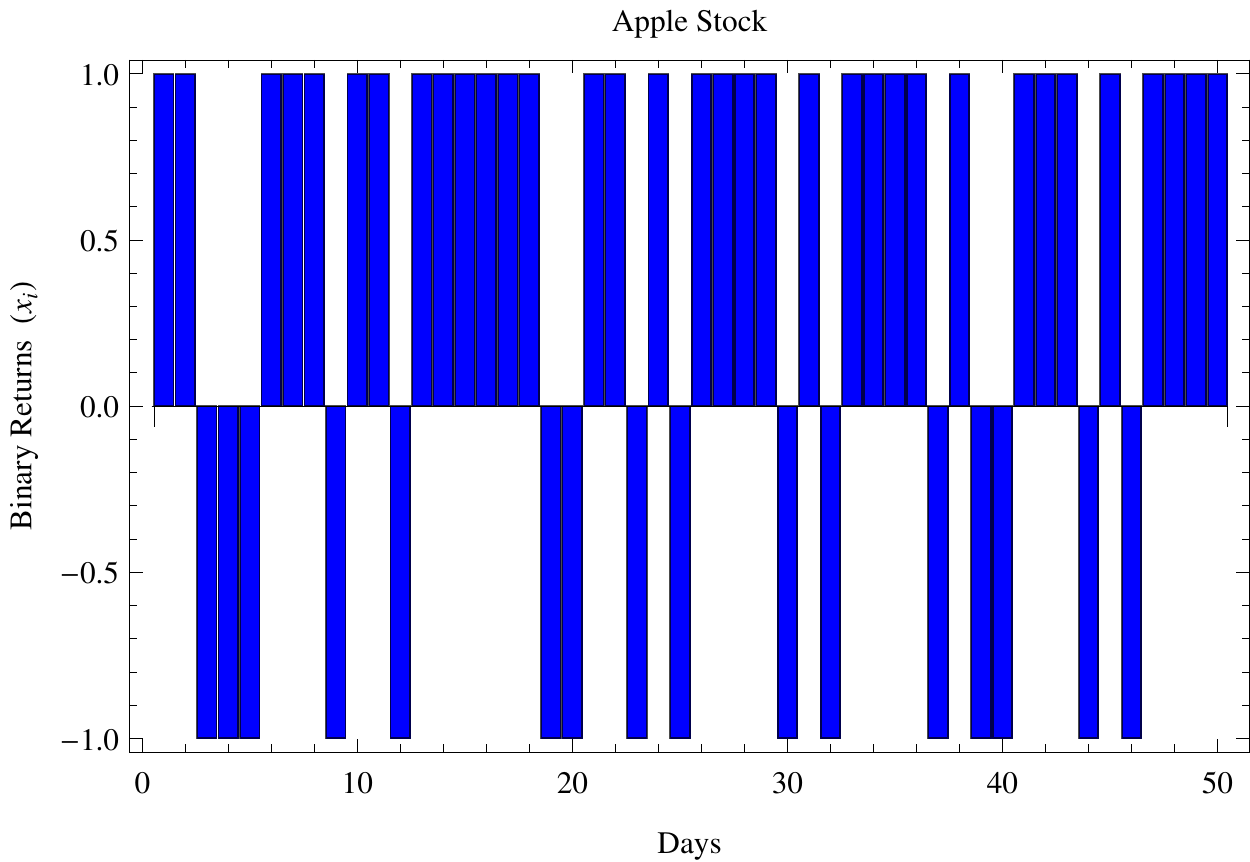}
\caption{`Weighted' (left) versus `binary' (right) time series of log-returns of the Apple stock over a period of 50 days starting from 7/5/2011.}
\label{fig:binarize}
\end{figure}
%%%%%%%%%%%%
 
We take logarithms of daily closing prices to obtain time series of \emph{log-prices} that represent our original `signal' $s_i(t)$, where $i$ labels stocks and $t$ labels days in the sample.
Correspondingly, we construct time series of \emph{log-returns} where each entry represents the increment $r_i(t)$ as defined in eq.(\ref{eq:r}).
Finally, we take the sign $x_i(t)$ of each
log-return $r_i(t)$ to obtain an additional, binarized set of time series as in eq.(\ref{eq:sign}).
We will refer to the binarized time series as the \emph{binary projection} of the original time series.
In fig. \ref{fig:binarize} we show a simple example of a weighted time series, along with the corresponding binary projection.
The (multiple) time series of $r_i(t)$ and $x_i(t)$ are the main objects of our analysis throughout the paper.
Note that, while the use of log-returns rather than simple returns (i.e. price differences) in finance is an important step that allows to remove overall trend effects over long time spans \cite{Mantegna(a)}, the binary signature is actually independent of whether the original prices have been logarithmically transformed.

%%%%%%%%%%%%%
\begin{figure*}[htb!]
\centerline{\includegraphics[width=.99\linewidth]{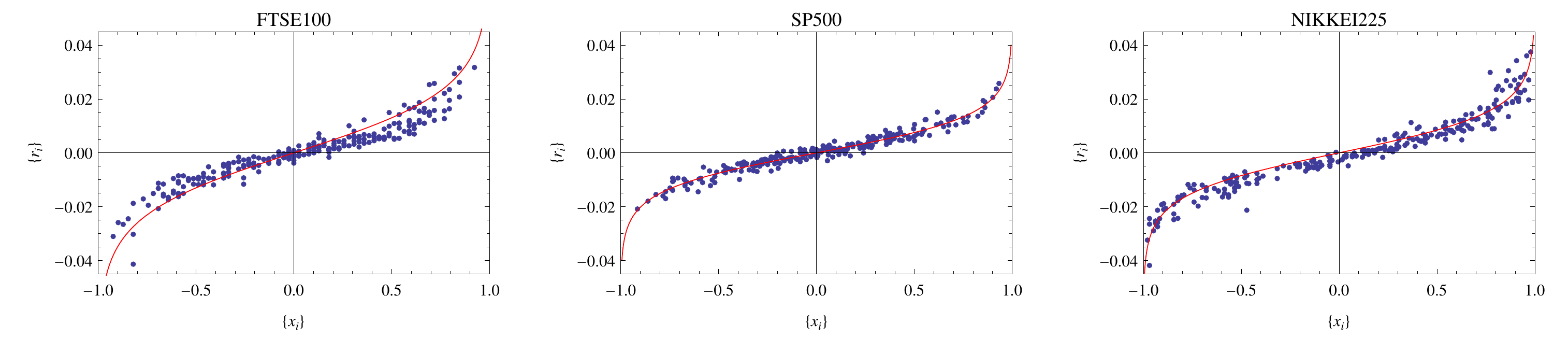}}
\caption{Nonlinear relationship between the average daily increment (weighted return) and the average daily sign (binary return) over all stocks in the FTSE100 (left), S\&P500 (center) and NIKKEI225 (right) in various years (2003, 2007, and 2004 respectively). 
Here each point corresponds to one day in the time interval of 250 trading days (approximately one year). The red line \color{red} \color{black} represents the best fit with the function $y=a\cdot\textrm{artanh}x$, whose use is theoretically justified later in sec.\ref{sec:full}.}
\label{fig:empir1}
\end{figure*}
%%%%%%%%%%%%

The main reason why for choosing the daily frequency is to achieve an optimal level of structural compatibility between the data and the models we introduce later. 
As we discuss in detail in sec. \ref{sec:methods}, our models are binary, i.e. they only allow the two values $\pm 1$ depending on whether the increment of the original time series is positive or negative. 
An increment of $0$ is not admitted in the models: consistently, we chose a frequency for which zero increments are extremely rare in the data. 
In financial markets, this is the case for daily (or lower) frequency. Indeed, a zero return value occurs in less than 0.2 \% of the cases in our daily data (when this happens, we randomly switch the corresponding binary increment to either $+1$ or $-1$ with equal probability). Higher-frequency data feature an increasing percentage of zero returns, a property that calls for an extension of the models considered here.

It should be noted that other types of binary time series, different from the $\pm 1$ type considered here, can also be defined.
Most notably, $0/1$ binary time series can indicate the occurrence of an event in a time period, i.e. whether the event happened ($1$) or not ($0$).
Financial examples include time series of  recession indicators \cite{Startz,Kauppi} or of `switching points' in stock returns \cite{switching}.
For such $0/1$ binary time series, correlations may not be very informative when measuring a dependence between the dichotomous variables.
To confront this gap, in recent years new methods were introduced, like the auto-persistence function and auto-persistence graph \cite{Startz}.  
In these methods, the dependence structure among the observations is described in terms of conditional probabilities, rather than correlations. 
Although throughout this paper we will be entirely focusing on $\pm1$ binary time series that naturally descend from the original \emph{signed} time series of fluctuations, it is interesting to notice that our approach can be extended, with slight modifications, to $0/1$ time series as well. 
To this end, one needs to re-express all quantities in terms of a $0/1$ binary variable $y\equiv (x+1)/2$, where $x$ is our $\pm 1$ binary variable, and adapt our approach accordingly.

\subsection{Nonlinear binary/non-binary relationships\label{sec:empirical}}
We now come to the main empirical findings that motivate our paper.
For each index and for each day $t$ in the sample, we first calculate the average (over all stocks) weighted return, that we denote as $\{r_i(t)\}$ and define as
\begin{equation}
\{r_i(t)\}\equiv\frac{1}{N}\sum_{i=1}^N r_i(t).
\label{eq:defr}
\end{equation}
Note that the above expression does not depend on the particular stock $i$, but it does depend on time $t$. Our unconventional choice of the symbol $\{\cdot\}$ to denote an average over stocks is to avoid confusion with temporal averages, that will be denoted by the more usual bar ($\overline{\cdot}$) later in the paper.
Similarly, we calculate the corresponding average binary return $\{x_i(t)\}$, defined as
\begin{equation}
\{x_i(t)\}\equiv\frac{1}{N}\sum_{i=1}^N x_i(t)
\label{eq:defx}
\end{equation}

In fig.\ref{fig:empir1} we plot $\{r_i(t)\}$ as a function of $\{x_i(t)\}$ for all days of various 1-year intervals and for the three indices separately.
We find a strong non-linear dependency between the two quantities.
Note that the average binary return is bound between $-1$ and $+1$ by construction, but the average weighted return is unbounded from both sides.
While there are in principle infinite values of $\{r_i(t)\}$ that are consistent with the same value of $\{x_i(t)\}$, we observe a tight relationships between the two quantities.
This relationship can be fitted by a one-parameter curve of the form 
\begin{equation}
\{r_i(t)\}=a\cdot\textrm{artanh}\big[\{x_i(t)\}\big]=\frac{a}{2}\ln\frac{1+x_i(t)}{1-x_i(t)}
\label{eq:artan}
\end{equation} 
(the theoretical justification for this functional form will be given in sec.\ref{sec:full}), where $a$ is in general different for different years and different indices.
Still, as we show later, for a given year and market the average weighted return of any day $t$ is to a large extent predictable (out of sample) from the average binary return of the same day, once $a$ is known (for instance by fitting the above curve to the data for a past time window).
In sec.\ref{sec:full} we will also show that the nonlinear character of the observed relations is a genuine signature of correlation in the data, as an uncorrelated null model shows a completely linear behaviour. 

%%%%%%%%%%%%%
\begin{figure*}[t]
\centerline{\includegraphics[width=.99\linewidth]{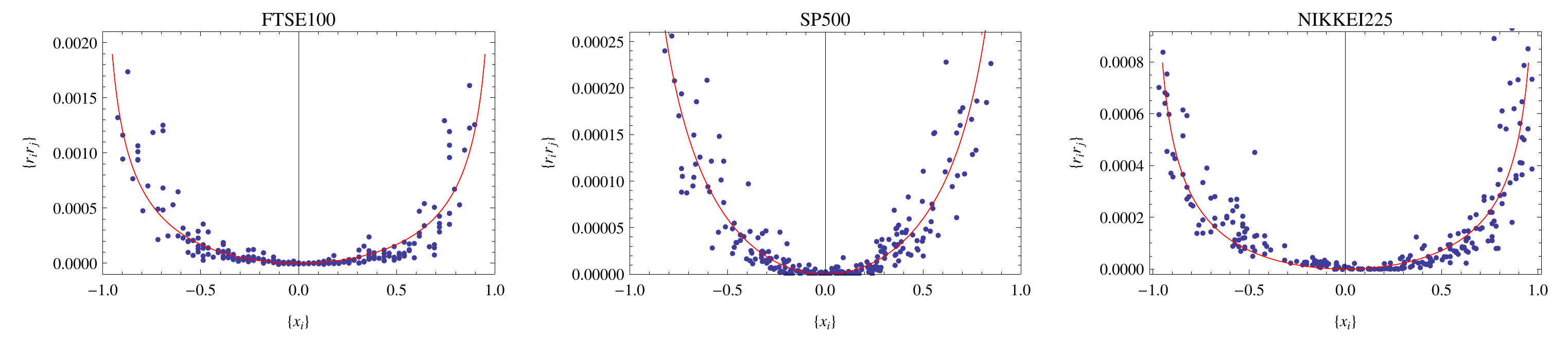}}
\caption{Nonlinear relationship between the average daily coupling (weighted coupling) and the average daily sign (binary return) over all stocks in the FTSE100 (left), S\&P500 (center) and NIKKEI225 (right) in various years (2003, 2007, and 2004 respectively). 
Here each point corresponds to one day in the time interval of 250 trading days (approximately one year). The red line represents the best fit with the function $y=b\cdot(\textrm{artanh}x)^2$, whose use is theoretically justified later in sec.\ref{sec:full}.}
\label{fig:empir2}
\end{figure*}
%%%%%%%%%%%%

There is another empirical relationship, involving a higher-order quantity. For each index and for each day $t$ in the sample, we calculated what we will call the average `coupling' over the $N(N-1)/2$ distinct pairs of stocks:
\begin{equation}
\{r_i (t)r_j(t)\}\equiv \frac{2\sum_{i<j}r_i(t) r_j(t)}{N (N-1)}
\label{eq:defcoupling}
\end{equation}
(so now the symbol $\{\cdot\}$ indicates an average over \emph{pairs} of stocks).
In fig.\ref{fig:empir2} we plot $\{r_i(t)r_j(t)\}$ as a function of the average binary return $\{x_i(t)\}$, for the same data as in fig. \ref{fig:empir1}.
Again, we find a strong non-linear dependency, where for a given value of the average binary return of day $t$ there is a typical value of the average coupling among all stocks in the same day. 
The relationship can be fitted by a one-parameter curve that diverges at $\{x_i\}=\pm 1$. As we show in sec.\ref{sec:full}, an uncorrelated null model would yield a different, parabolic curve with no divergences. 
Again, this means that the empirical trend is due to genuine correlations, whose nature will be clarified later on in the paper.

There are even more examples of dependencies that we can find between binary and non-binary properties in the data. However, in one way or another all these relationships, including that shown in fig.\ref{fig:empir2}, ultimately derive from eq.(\ref{eq:artan}). 
For this reason, we refrain from showing redundant results and focus on the empirical findings discussed so far.

The above analysis indicates that the binary signature of financial time series contains  relevant information about the original data. 
While the binary signature is \emph{a priori} a many-to-one projection involving a significant information loss, we empirically find that there are properties (namely the average return and average coupling) for which the projection is virtually a one-to-one `quasi-stationary' transformation (on appropriate time scales, as we show in sec.\ref{sec:full}), allowing to reconstruct the corresponding original, weighted properties to a great extent.
Rather than exploring the practical aspects of this possibility of reconstruction of the original signal from its binary projection, in this paper we are interested in understanding the origin of this behaviour and providing a simple data-driven model of it.
This will be ultimately achieved in sec.\ref{sec:full}, where we also show that the binary/non-binary relations we have documented are a novel quantification of the fact that extreme price increments occur more often when most stocks move in the same direction.
This is an important type of correlation between the magnitude of log-returns of individual time series and the level of synchronization (common sign) of the increments of all stocks in the market.

\section{Maximum-entropy matrix ensembles\label{sec:methods}}
Having established that the binary projections of real time series contain non-trivial information, in the rest of the paper we introduce a theory of binary time series aimed, among other things, at reproducing the observed non-linear relationships showed in figs. \ref{fig:empir1} and \ref{fig:empir2}.
In our approach, we regard a synchronous set of binary time series as a $\pm 1$ matrix and we introduce an ensemble of such matrices via the maximization of Shannon's entropy, subject to the constraint that some specified properties of the ensemble match their observed values.
An analogous approach is widely used e.g. in network analysis and known under the name of Exponential Random Graphs \cite{Park J and Newman}. 
Moreover, we provide an analytical maximum-likelihood method to find the optimal values of the paramaters governing the ensembles, which is again similar in spirit to a method that has been recently introduced for networks \cite{Garlaschelli1,Garlaschelli2,Garlaschelli3}.
Finally, we describe Akaike's information criterion (AIC) \cite{Akaike74}, which we will use to rank and compare the performance of different null models when fitted to the same data. 

Being entropy-based, our approach automatically allows us to measure the amount of information encoded into the observed properties chosen as constraints, i.e. how much information is gained about the original (set of) time series once those properties are measured.
It also allows us to identify, given a set of measured properties, which ones are more informative and which ones can be discarded, as we show on specific financial examples.
Our framework turns out to reproduce the observed non-linear relationships very well, thus providing a simple mathematical explanation and functional form for the plots shown in the previous section.
Moreover, we are able to identify, as a function of the binary return only, distinct regimes in the collective behavior of stocks, namely a `coordinated' regime dominated by market-wide interactions, an `uncoordinated' regime dominated by stock-specific noise and an `intermediate' regime where both market-wide and stock-specific information is relevant. 

We incidentally note that, despite the available variety of refined and advanced techniques in time series analysis \cite{chaos}, how one can quantify (in the sense of statistical ensembles) how much information is actually encoded into any given, measurable property of a time series is still not fully understood.
While most studies, starting from the celebrated work by Kolmogorov about the algorithmic complexity of sequences of symbols \cite{kolmogorov}, have addressed the quantification of the information content of a single time series, much less is known about the information encoded in the measured value of a given time series property (which, necessarily, involves the idea of an entire \emph{ensemble} of time series consistent with the measured value itself).  
Our approach can provide an answer to such a question, by associating an absolute
level of uncertainty (entropy) to each observable of an empirical (set of) time series. 
In relative terms, this also allows us to compare the information content of different properties of a time series, thereby indicating which measured property is the most informative about the original time series. 

As a final consideration, it is worth mentioning that the maximum-entropy matrix ensembles that we introduce are clearly related to (and, depending on the specification, potentially overlapping with) some ensembles that are well studied by random matrix theory \cite{Laloux, Plerou, Plerou1, Plerou2, Gopikrishnan, Rosenow}.
However, our approach is different since we generate ensembles of matrices whose probability distributions are determined by the kind of partial information (empirically measured constraint) about the real system. 
In this approach the maximization of Shannon's entropy, given some real-world available information, yields the least biased probability distribution (over the space of possible matrices) consistent with the data. This formalism allows us to relate the probabilistic structure of each matrix ensemble with the choice of the original observed property, or constraint. 
Similarly, since our matrices represent (multiple) time series, we are able to connect the various ensembles to simple stochastic processes induced by the associated matrix probabilities and, again, to the chosen empirical property specifying the ensembles themselves.

\subsection{Exponential Random Matrices\label{sec:ERM}}
We first analytically characterize the properties of families of randomized matrices. More generally, we introduce a matrix ensemble that maximizes Shannon's entropy, while enforcing a set of observed constraints (selected time series properties). This procedure is analogous to e.g. that leading to the definition of Exponential Random Graphs in network theory \cite{Park J and Newman}. 
However, we will modify it to accommodate $\pm  1$ matrices, as opposed to $0/1$ or non-negative matrices that describe binary and weighted networks respectively.
The resulting ensemble can thus be denoted as the `Maximum-Entropy Matrix' (MEM) ensemble or equivalently `Exponential Random Matrices' (ERMs) model. 

Let us consider the ensemble of all $\pm 1$ matrices with dimensions $N\times T$.
Each such matrix can represent $N$ synchronous time series, all of duration $T$ (for instance, if applied to a set of multiple financial time series, $N$ refers to the number of stocks and $T$ to the number of time steps). 
Let $\mathbf{X}$ denote a generic matrix in the ensemble, and $x_i(t)$ its entry ($1\le i\le N$, $1\le t\le T$).
Let $\mathbf{X} ^*$ be the particular real matrix that we observe. In other words, our ensemble is composed of all possible matrices $\mathbf{X}$ of the same type as $\mathbf{X} ^*$, and includes $\mathbf{X} ^*$ itself. 
For any data-dependent property $R$, we will consider the value $R(\mathbf{X})$ obtained when $R$ is measured on the particular matrix $\mathbf{X}$. 
For each matrix $\mathbf{X}$ in the ensemble, we will assign an occurrence probability $P(\mathbf{X})$. 
The expectation value (ensemble average) of a
property $R$ can be expressed as
\begin{equation}
\langle R \rangle=\sum_\mathbf{X} R(\mathbf{X})P(\mathbf{X})
\label{eq:R}
\end{equation}
where the sum runs over all matrices in the ensemble. 

At this point, we introduce a set of constraints denoted by the vector $\vec{C}$. 
The constraints are meant to ensure that a given set of observed properties $\vec{C}(\mathbf{X}^*) $ in the real matrix $\mathbf{X} ^*$ is reproduced by the ensemble itself. 
In our method we will enforce `soft' constraints by requiring that their expectation value  $\langle\vec{C}\rangle$ equals the observed one. The resulting ensemble is a \emph{canonical} one where each matrix $\mathbf{X}$  is assigned a probability $P(\mathbf{X})$ that maximizes Shannon's entropy
 \begin{equation}
S\equiv-\sum_\mathbf{X} P(\mathbf{X})\ln{P(\mathbf{X})}
\end{equation}
subject to the normalization constraint
\begin{equation}
\sum_\mathbf{X} P(\mathbf{X}) =1
\label{eq:P}
\end{equation}
and to the chosen vector of constraints 
\begin{equation}
\langle\vec{C}\rangle=\sum_\mathbf{X} C(\mathbf{X})P(\mathbf{X})=\vec{C}
\label{eq:C}
\end{equation}
that we are enforced in order to reproduce the desired set of observed quantities. 

The solution to the above constrained maximization problem is standard (see for instance \cite{Park J and Newman} for a recent derivation in the context of networks). We first introduce the Lagrange multipliers $\alpha$ and $\vec{\theta}$, enforcing eqs.\eqref{eq:P} and \eqref{eq:C} respectively, and then require that the functional derivative of Shannon's entropy (plus the constraining terms) vanishes: 
\begin{eqnarray}
\frac {\partial}{\partial P(\mathbf{X})} \Big\{S+\alpha\Big[1- \sum_\mathbf{X} P(\mathbf{X})\Big] +} \displaystyle{\sum_i \theta_i\Big[C_i-\sum_\mathbf{X} C(\mathbf{X})P(\mathbf{X})\Big]\Big\}=0
\nonumber
\end{eqnarray}
This yields
\begin{equation}
\ln{P(\mathbf{X})} +1 +\alpha +\sum_i \theta_i C_i(\mathbf{X}) =0
\end{equation}
for any matrix $\mathbf{X}$. 
Using a notation that makes the dependence of all quantities on ${\vec{\theta}}$ explicit, we then obtain
\begin{equation}
P(\mathbf{X}|\vec{\theta})=\frac{e^{-H(\mathbf{X},\vec{\theta})}}{Z(\vec{\theta})}
\label{eq:PX}
\end{equation}
where $H(\mathbf{X},\vec{\theta})$ is the \emph{Hamiltonian} 
\begin{equation}
H(\mathbf{X},\vec{\theta})\equiv\vec{\theta}\cdot\vec{C}(\mathbf{X})=\sum_i \theta_i C_i(\mathbf{X}),
\label{eq:HX}
\end{equation}
which is a linear combination of the constraints, and $Z(\vec{\theta})$ is the \emph{partition function}
\begin{equation}
Z(\vec{\theta}) \equiv e^{\alpha+1} = \sum_\mathbf{X} e^{-H(\mathbf{X},\vec{\theta})},
\label{eq:ZX}
\end{equation}
which is the normalizing constant for the probability. 
Consistently, we can rewrite eq. \eqref{eq:R} more explicitly as a function of $\vec{\theta}$:
\begin{equation}
\langle R \rangle_{\vec{\theta}}\equiv \sum_\mathbf{X} R(\mathbf{X})P(\mathbf{X}|\vec{\theta})
\end{equation}
where $\langle\cdot\rangle_{\vec{\theta}}$  indicates that the ensemble average is evaluated at the particular parameter value ${\vec{\theta}}$.

Equations \eqref{eq:PX} to \eqref{eq:ZX} define the MEM or ERM model. Specifically, the model yields the probability distribution over a specified ensemble of matrices, which maximizes the entropy under a set of generic constraints. 
The guiding principle is that the probability distribution (over microscopic states) which have maximum entropy, subject to observed (macroscopic) properties, provides the most unbiased representation of our knowledge of the state of a system\cite{Jaynes2}. 
To put it in a more physical frame, this is analogous to the Gibbs-Boltzmann distribution over the microstates of a large system at a well defined temperature, given the thermodynamic (macroscopic) observables such as the total energy. 

\subsection{Maximum-likelihood parameter estimation\label{sec:ML}}
The above derivation shows that the expectation value of any property of the ensemble depends \emph{functionally} on the specific enforced constraints $\vec{C}$ through the resulting structure of $P(\mathbf{X}|{\vec{\theta}})$. 
Of course, it also depends \emph{numerically} on the measured values $\vec{C}(\mathbf{X}^*)$ of the constraints themselves, through the particular parameter value (that we denote by $\vec{\theta}^*$) required in order to enforce that the expected and observed values of $\vec{C}$ match:
\begin{equation}
\langle \vec{C}\rangle_{\vec{\theta}^*}=\vec{C}(\mathbf{X}^*).
\label{eq:match}
\end{equation}

We now show that the value $\vec{\theta}^*$ that satisfies eq.\eqref{eq:match} concides with the value that maximizes the likelihood to generate the empirical data, as in the corresponding Maximum Likelihood (ML) approach to network ensembles \cite{Maximum likelihood,Garlaschelli1}. 

We start by writing the log-likelihood function of an observed matrix  $\mathbf{X}^*$  generated by the parameters $\vec{\theta}$:
\begin {equation}
 \lambda (\vec{\theta}) \equiv \ln{ P(\mathbf{X}^*|\vec{\theta}) }=-H(\mathbf{X}^*,\vec{\theta}) -\ln{Z(\vec{\theta})}
\end {equation}
We then look for the particular value $\vec{\theta}^*$ that maximizes $\lambda (\vec{\theta})$, i.e.
\begin {equation}
\vec {\nabla} \lambda (\vec{\theta}^*) = \left[  \frac{\partial \lambda (\vec{\theta})}{\partial \vec{\theta}}  \right]_{\vec{\theta}=\vec{\theta}^*}=\vec 0
\end {equation}
(it is easy to check that the higher-order derivative confirm that $\vec{\theta}^*$ is a point of maximum). This leads to  
\begin {equation}
\vec {\nabla} \lambda (\vec{\theta}^*) =  \left[ -\vec{C}(\mathbf{X}^*) - \frac{1}{Z (\vec{\theta})}  \frac{\partial  Z(\vec{\theta})}{\partial \vec{\theta}}   \right]_{\vec{\theta}=\vec{\theta}^*}=\vec 0
\end {equation}
the solution for that yields the ML condition
\begin {equation}
\vec{C}(\mathbf{X}^*) = \sum_\mathbf{X} \frac{\vec{C}(\mathbf{X}) e^{-H(\mathbf{X},\vec{\theta}^*)}}{Z(\vec{\theta}^*)} = \langle\vec{C}\rangle_{\vec{\theta}^*}.
\label{eq:match2}
\end {equation}
which coincides with eq.\eqref{eq:match}.
Thus the likelihood of the real matrix $\mathbf{X}^*$ is maximized by the specific parameter choice such that the ensemble average of each constraint equals its empirical value measured on $\mathbf{X}^*$ , automatically ensuring that the desired constraints are met.

\subsection{Model selection\label{sec:AIC}}
We finally show how we can use Akaike's information criterion (AIC) to rank the performance of different models, i.e. different choices of the constraints, in reproducing the same data. 
The AIC is an information-theoretic measure of the relative goodness of fit of a model, as compared to a set of alternative models all used to explain the same data \cite{Akaike74}. It offers a relative measure of the information lost when the given model is used to describe reality. 
The power of AIC (and other similar criteria \cite{Burnham and Anderson 2002}) lies in the possibility to rank a set of models in terms of their achieved trade-off between accuracy (good fit to the data) and parsimony (low number of free parameters) \cite{Burnham and Anderson 2002}. 
In general, for the $k$-th model in a set of selected models, AIC is defined as 
\begin{equation}
\textrm{AIC}_k=2n_k-2\lambda^*_k
\end{equation}
where $n_k$ is the number of free parameters in the $k$-th model and $\lambda^*_k$ is the maximized log-likelihood of the data under the same model.
The above expression effectively discounts the number $n_k$ of parameters (complexity) from the maximized likelihood $\lambda^*_k$ (accuracy).
The model with the lowest value of $\textrm{AIC}_k$ (let us denote this value by $\textrm{AIC}_{min}$) is the `best' model in the considered set, achieving the optimal trade-off  \cite{Burnham and Anderson 2002}.

In the ERM/MEM family of models we have introduced, a model is uniquely specified by the choice of the constraints $\vec{C}$.
Given a $N\times T$ data matrix $\mathbf{X}^*$ and a set $\{\vec{C}_1,\dots,\vec{C}_m\}$ of $m$ possible choices of constraints, each of the resulting $m$ models has an AIC value
\begin{equation}
\textrm{AIC}_k=2n_k-2\ln{ P_k(\mathbf{X}^*|\vec{\theta}^*_k) }\qquad k=1,m
\label{eq:AIC}
\end{equation}
where $n_k$ is the dimensionality of the vector $\vec{C}_k$, $\ln{ P_k(\mathbf{X}^*|\vec{\theta}^*_k) }$ is the maximized log-likelihood of model $k$, and $\vec{\theta}^*_k$ is the parameter value maximizing such log-likelihood.
Within our framework, AIC identifies which measured property $\vec{C}_k(\mathbf{X}^*)$ is most informative about the entire time series $\mathbf{X}^*$.

In order to understand whether models with values of AIC larger than but close to $\textrm{AIC}_{min}$ are still competitive, it is customary to define the so-called `AIC weights' which provide a normalized strength of evidence for a model  \cite{Burnham and Anderson 2002}.
For each model $k$ in the set of $m$ models, one first calculates the difference $\Delta_k = AIC_k - AIC_{min}$ and then defines the AIC weight  
\begin{equation}
w_k \equiv\frac{e^{{-\Delta_k}/{2}}}{\sum_{r=1}^{m}e^{{-\Delta_r}/{2}}}.
\label{eq:AICw}
\end{equation}
The AIC weight $w_k$ represents the probability that the $k$-th model is the best one among the $m$ selected models.
For instance, an AIC weight of $w_k=0.75$ indicates that, given the data, model $k$ has a $75\%$ chance of being the best model among the $m$ candidate ones.
If two or more models have comparable AIC weights (e.g. $w_1=0.6$, $w_2=0.4$ or $w_1=0.35$, $w_2=0.25$, $w_3=0.4$), then there is no evidence that the model with the highest AIC weight (lowest AIC value) is clearly outperforming the other ones. 
All the models with comparable weights should be considered as competing alternatives, in principle leading to the problem of multi-model inference \cite{Burnham and Anderson 2002}.

\section{Single time series\label{sec:sts}}
In this section we consider the first family of specifications of our general approach outlined in sec.\ref{sec:methods}.
We focus on the simple case of single time series ($N=1$), where the ensemble of $N\times T$ matrices reduces to an ensemble of $1\times T$ matrices, or equivalently of $T$-dimensional row vectors. Each such vector will still be denoted by $\mathbf{X}$. We assume long time series, i.e. $T\gg 1$.

This first specification of our abstract formalism is not meant to provide realistic models for the evolution of the binary increments of real financial time series. Rather, it allows us to make different sorts of considerations.
On one hand, it allows us to introduce our formalism using simpler examples first, establishing the basis for the more general cases (leading to the main results of the paper) that will be introduced later.
On the other hand, it emphasizes that different and well known (one-dimensional) stochastic processes are found as particular examples of maximum-entropy ensembles defined by specific constraints that are otherwise obscure. 
Identifying these `driving constraints' underlying common stochastic processes will help us interpret such processes in the light of the empirical properties being reproduced. 
Finally, our approach allows us to identify, given the data and given a set of simple properties, which of these properties is encoding the largest amount of information about the original binary signature.

Let $\mathbf{X}$ denote a single time series with entries $x(t)$, where $1\le t\le T$, each representing a temporal increment. 
We will denote the average increment (first moment) as
\begin{equation}
M_1(\mathbf{X})\equiv\overline{x(t)}=\frac{1}{T}\sum^T_{t=1} x(t).
\end{equation}
Note that the second moment is always
\begin{equation}
M_2(\mathbf{X})\equiv\overline{x^2(t)}=\frac{1}{T}\sum^T_{t=1} x^2(t)=1,
\end{equation}
so the sample variance is
\begin{equation}
M_2(\mathbf{X})-M_1^2(\mathbf{X})=1-\overline{x(t)}^2.
\end{equation}
We also define the $\tau$-delayed product (with $0\le\tau\le T$)
\begin{equation}
B_\tau(\mathbf{X})\equiv\overline{x(t)\cdot x(t+\tau)}=\frac{1}{T}\sum_{t=1}^T x(t)\cdot x(t+\tau)
\label{eq:B}
\end{equation}
where we have introduced periodic boundary conditions:
\begin{equation}
x(T+\tau)\equiv x(\tau)\quad\textrm{with}\quad 0\le\tau\le T
\label{eq:mod}
\end{equation}
The above periodicity condition in inessential, since we could have used a definition avoiding its introduction, but it makes some expressions simpler in what follows.
Periodicity implies that the normalized (between $-1$ and $+1$) autocorrelation function (with delay $\tau$) can be defined as
\begin{eqnarray}
A_\tau(\mathbf{X})&\equiv&\frac{\overline{x(t)\cdot x(t+\tau)}-\overline{x(t)}\cdot \overline{ x(t+\tau)}}{{\overline{x^2(t)}-\overline{x(t)}^2}}\nonumber\\
&=&\frac{B_\tau(\mathbf{X})-M_1^2(\mathbf{X})}{1-M_1^2(\mathbf{X})}
\label{eq:autocorrdef}
\end{eqnarray}

\begin{figure*}[t]
\includegraphics[scale=0.55]{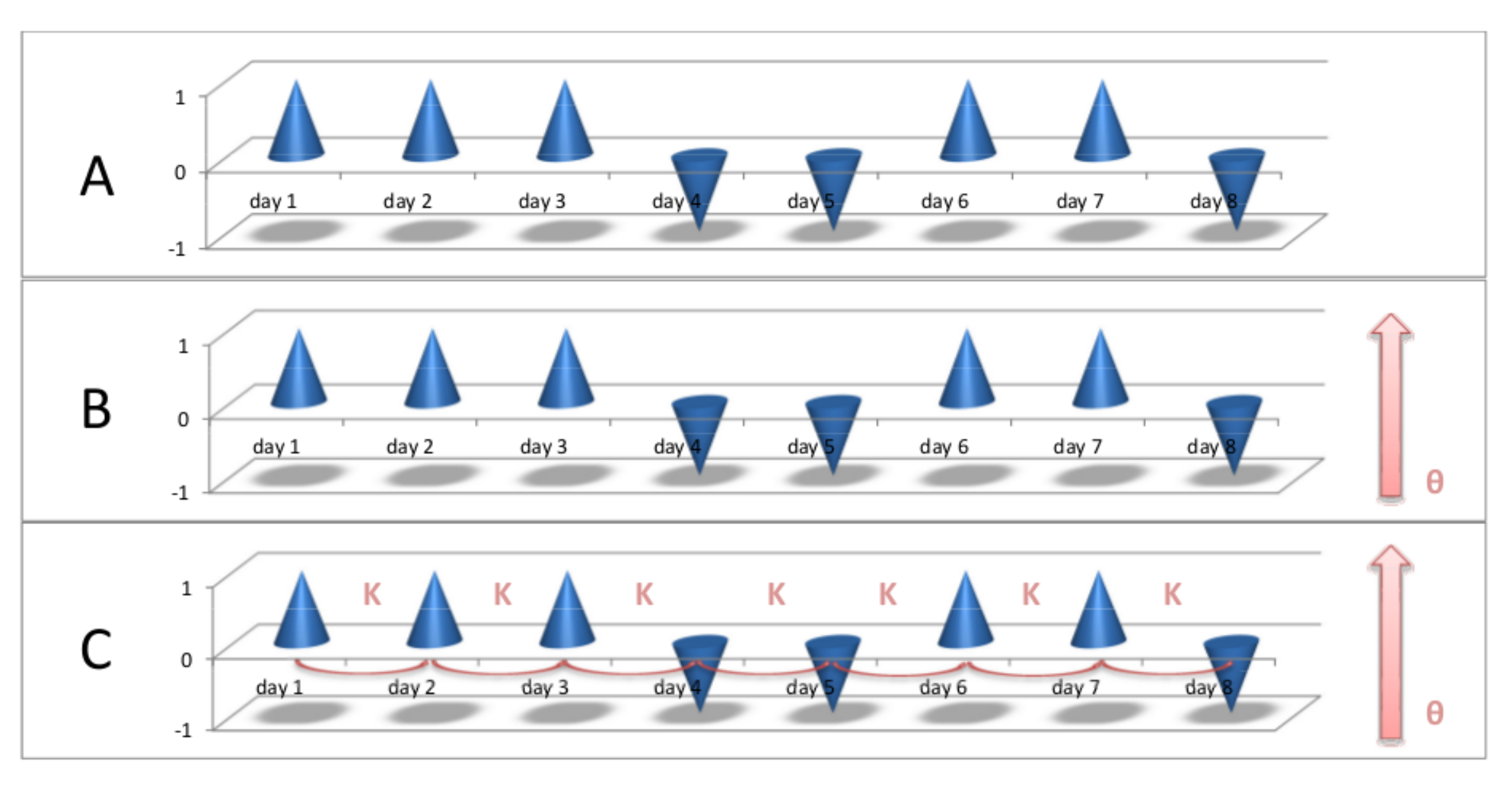}
\caption{Illustration of our mapping from single binary time series to spin models.
Each time series is regarded as a chain of $\pm 1$ spins, where the value of the spin indicates if the daily return of the stock is positive ($+1$) or negative ($-1$) . 
In each model we enforce different constraints, that imply different spin models and different stochastic processes. 
Given the same time series, we consider three possible models. A) we enforce no constraint, which translates into a chain of non-interacting spins without external field (uniform random walk).
B) we enforce the total temporal increment, which translates into a chain of non-interacting spins with external field (biased random walk).
C) we enforce both the total increment and the one-lagged autocorrelation, which translates into a chain of spins with first-neighbour interactions and external field (markov process).}
\label{fig:singlespins}
\end{figure*}

Since a ($\pm 1$) binary time series can also be regarded as a chain of classical spins pointing either up or down, it is straightforward to consider simple, analytically solved spin models as the starting point, since these models are defined in terms of a `physical' Hamiltonian that has precisely the same structure of our `information-theoretic' Hamiltonian defined in eq.\eqref{eq:HX}.
In what follows, we introduce various model specifications. 
For each model, we introduce the constraints that we enforce and the resulting Hamiltonian as described in sec.\ref{sec:ERM}. Different constraints correspond to different spin models and lead to different stochastic processes. This is pictorially illustrated in fig. \ref{fig:singlespins}. 
The free parameters conjugated to the constraints will be fitted according to the Maximum Likelihood principle described in sec.\ref{sec:ML}. Different models will be ranked according to the AIC weights introduced in sec.\ref{sec:AIC}.

%%%%%%%%%%%%
%%%%%%%%%%%%

\subsection{Uniform random walk\label{sec:singleurw}}
The most trivial model is one where we enforce no constraint, i.e. there is no free parameter and the Hamiltonian is
\begin{equation}
H(\mathbf{X})=0.
\end{equation}
Physically, the above Hamiltonian describes a gas of $T$ non-interacting `spins' in vacuum, i.e. in absence of an external magnetic field.
This model is discussed in the Appendix. The probability of occurrence  of a time series $\mathbf{X}$ is completely uniform over the ensemble of all binary time series of length $T$. 
All the $T$ elements of $\mathbf{X}$ are mutually independent and identically distributed. This results in a completely uniform random walk with zero expected value for each increment:
\begin{equation}
\langle x(t)\rangle=0 
\end{equation}
While the (ensemble) variance of each increment equals
\begin{equation}
\textrm{Var}[x(t)]\equiv\langle x^2(t)\rangle-{\langle x(t)\rangle}^2=1.
\end{equation}

This trivial model generates a symmetric random walk. Since the expected return is zero, and the uncertainty is maximal, the variance is also maximal (for a $\pm1$ binary random variable). 
Financially, the model assumes that the stock fluctuates randomly, with no memory, and with no overall `price drift'. This is the most basic model of price dynamics that has been considered in the financial literature since the pioneering work of Bachelier \cite{Mantegna&Stanley}, here adapted to the case of binary time series.

The model can be used as a basic benchmark for checking the performance of our other models. This comparison will be studied in sec. \ref{sec:comparing}.
Since here the likelihood is independent of any parameter, the AIC of the model can be calculated using eq.\eqref{eq:AIC} where the probability is given by eq.\eqref{eq:appPrw} (see Appendix) and the number of parameters is $n_k=0$. 

\subsection{Biased random walk\label{sec:singlebrw}}
We now consider the total increment as the simplest non-trivial (one-dimensional) constraint:
\begin{equation}
C(\mathbf{X})=T\cdot M_1(\mathbf{X})=T\cdot\overline{x(t)}
\label{eq:Cbiased}
\end{equation} 
This leads to the Hamiltonian
\begin{equation}
H\left(\mathbf{X},\theta\right)=\theta \sum_{t=1}^T x(t),
\end{equation}
which coincides with the physical Hamiltonian for a gas of $T$ non-interacting `spins' in a common external `magnetic field' $-\theta$.

As we show in the Appendix, this model generates a \emph{biased} random walk where the probability $P_t(x|\theta)$  of a given increment $x=\pm 1$ at time $t$ is
\begin{equation}
P_t(x|\theta)= \frac{ e^{-\theta x}}{ e^{-\theta} +e^{+\theta}}.
\end{equation}
The expected return is the hyperbolic tangent
\begin{equation}
\langle x(t)\rangle_\theta=-\tanh{\theta},
\end{equation}
while the variance is 
\begin{equation}
\textrm{Var}[x(t)]=1- \tanh^2{\theta}.
\end{equation} 
Financially, this model still assumes no memory in the fluctuations of a given stock, but it introduces a `price drift' in terms of a non-zero expected return.

The maximum likelihood condition \eqref{eq:match}, fixing the value $\theta^*$ of the parameter $\theta$ given a real time series $\mathbf{X}^*$, leads to
\begin{equation} 
\theta^*=-\frac{1}{2}\ln\left[\frac{1+\overline{x^*(t)}}{1-\overline{x^*(t)}}\right]
\label{eBRW}.
\end{equation}
The maximized likelihood for the model is
\begin{equation}
P(\mathbf{X}^*| \theta^{*}) = \prod_{t=1}^T P_t\big(x^*(t)|\theta^{*}\big)
\end{equation}
which, using eq.\eqref{eq:AIC} with $n_k=1$, can be used to measure the AIC (see sec.~\ref{sec:AIC}) of the model, based on the observed data.
This will be done in sec. \ref{sec:comparing}.

\subsection{One-lagged model\label{sec:singleolm}}
Let us now explore a more complex model of collective behavior. The models considered so far were non-interacting, i.e. each return in the time series was independent of the previous outcomes. 
Now we consider a model where, besides the constraint on the total increment specified in eq.(\ref{eq:Cbiased}), we enforce an additional constraint on the time-delayed (lagged) quantity $T\cdot B_1(\mathbf{X})$, where $B_1(\mathbf{X})$ is defined in eq.\eqref{eq:B} with $\tau=1$.
Financially, this amounts to enforce the average return \emph{and} the average one-step temporal autocorrelation of the time series. In order words, besides a price drift, we also introduce a short-term memory.

The resulting 2-dimensional constraint can be written as
\begin{equation}
\vec{C}(\mathbf{X})=\left(\begin{array}{c}C_1(\mathbf{X})\\C_2(\mathbf{X})\end{array}\right)=
T\cdot\left(\begin{array}{c} M_1(\textbf{X})\\ B_1(\textbf{X})\end{array}\right).
\end{equation}
If we write the corresponding Lagrange multiplier as
\begin{equation}
\vec{\theta}=\left(\begin{array}{c}\theta_1\\\theta_2\end{array}\right)=-
\left(\begin{array}{c} I\\K\end{array}\right),
\end{equation}
then the Hamiltonian reads
\begin{eqnarray} 
H(\textbf{X},I,K) &=&-I \sum_{t=1}^T x(t)
 -K \sum^{T}_{t=1} x(t)x(t\!+\!1), \end{eqnarray}
where we consider a periodicity condition as in eq.\eqref{eq:mod} with $\tau=1$, i.e. $x(T+1)\equiv x(1)$. Note that, when $\mathbf{X}$ is a real binary time series of length $T$, this condition can be always enforced by adding one last (fictious) timestep $T+1$ and a corresponding increment $x(T+1)$ chosen equal to $x(1)$.
For long time series (large $T$), the effects induced by this addition are negligible.

The above Hamiltonian coincides with that for the one-dimensional Ising model with periodic boundary conditions \cite{baxter}, which is a model of interacting spins under the influence of an external `magnetic' field $I$.
The model is analytically solvable (see Appendix for the complete derivation), which allows us to apply it to real time series in our formalism. 
In our setting, each time step $t$ is seen as a site in an ordered chain of length $T$, and each value $x(t)=\pm 1$ is seen as the value of a spin sitting at that site. `First-neighbour interactions' along the chain of spins are here interpreted as one-lagged memory effects.
As a result of these interactions, the model generates time series according to a Markov process where the probability of an increment $x(t+1)$ depends on the realized increment $x(t)$ at the previous time step $t$. This is evident from the solution of the model, see e.g. eq. \eqref{eq:twostep} in the Appendix.

The solution of the model yields the following expectation values
\begin{eqnarray}
\big\langle M_1\big\rangle_{I,K}&=& \frac{e^{2K}\sinh{I}}{\sqrt{1+e^{4K}\sinh^2{I}}}\\
\big\langle B_\tau\big\rangle_{I,K}&=& \frac{e^{4K}\sinh^2{I}+(\lambda_1/\lambda_2)^\tau}{1+e^{4K}\sinh^2{I}}
\end{eqnarray}
(see Appendix) where
\begin{eqnarray}
\lambda_1 &=& e^K \cosh I+\sqrt{e^{2K}\sinh^2 I+e^{-2K}},\\
\lambda_2 &=& e^K \cosh I-\sqrt{e^{2K}\sinh^2 I+e^{-2K}}.
\end{eqnarray}
The resulting expected value of the normalized autocorrelation defined in eq. \eqref{eq:autocorr} is simply
\begin{equation}
\big\langle A_\tau\big\rangle_{I,K}
= \left( \frac{\lambda_1}{\lambda_2}\right)^\tau.
\label{eq:autocorr}
\end{equation}

%%%%%%%%%%%%
\begin{figure*}[t]
\centering
\includegraphics[scale=0.55]{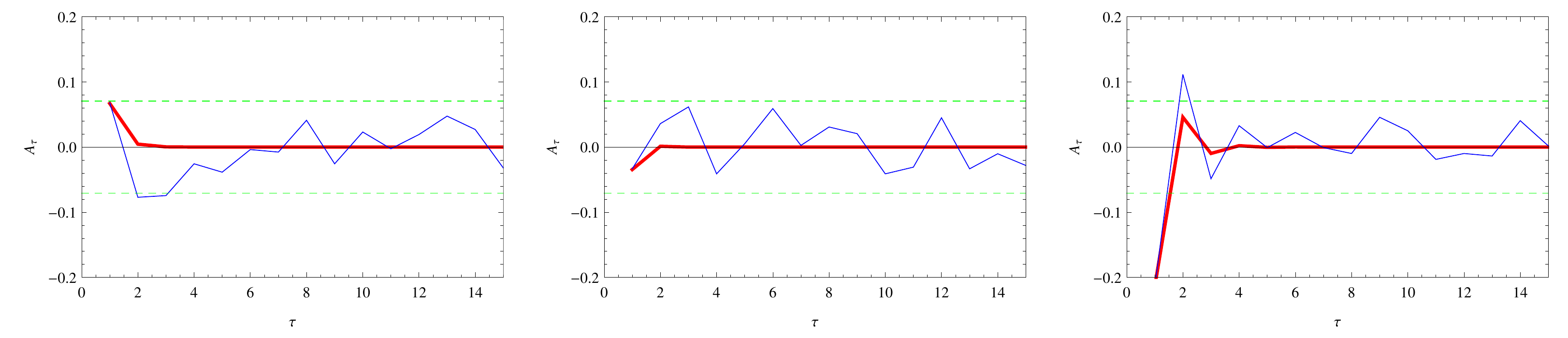}
\caption{Measured autocorrelation (blue) of three different S\&P500 stocks (Qcom, USB, and MJN respectively) over a period of 800 trading days (approximately 3.5 years), and comparison with the predicted autocorrelation $\langle A_\tau\rangle_{I,K}$ generated by the one-lagged (one-dimensional Ising) model (red). The green lines represent the noise level, calculated as $\pm 2$ standard deviations of the Fisher-transformed autocorrelation.}
\label{fig-5}
\end{figure*}
%%%%%%%%%%%%

The above expressions allow us to calculate all the relevant expected properties of the time series generated by the model, once the parameters $I$ and $K$ are set to the values $I^*$ and $K^*$ maximizing the likelihood $P(\mathbf{X}^*|I,K)$ of the observed time series $\mathbf{X}^*$. 
These values are the solutions of the coupled equations
\begin{eqnarray}
M_1(\mathbf{X}^*)&=&\big\langle M_1\big\rangle_{I^*,K^*}\label{eq:M1}\\
B_1(\mathbf{X}^*)&=&\big\langle B_1\big\rangle_{I^*,K^*}.\label{eq:B1}
\end{eqnarray}
where $M_1(\mathbf{X}^*)$ and $B_1(\mathbf{X}^*)$ are the empirical values measured on the real data $\mathbf{X}^*$.
The maximized likelihood of the model can be calculated as $P(\mathbf{X}^*|I^*,K^*)$, where $P(\mathbf{X}|I,K)$ is given by eq.~\eqref{eq:twostep} in the Appendix.
From the maximized likelihood, the AIC can be easily obtained using eq.\eqref{eq:AIC} with $n_k=2$.

Note that the values $I^*$ and $K^*$ are such that the first point of the expected autocorrelation function, $\langle A_1\rangle_{I^*,K^*}$, is necessarily equal to the observed value $A_1(\mathbf{X}^*)$.
Based on this first value alone, the model will provide the full expected autocorrelation $\langle A_\tau\rangle_{I^*,K^*}$ as follows:
\begin{equation}
\big\langle A_\tau\big\rangle_{I^*,K^*}
= \left( \frac{\lambda_1}{\lambda_2}\right)^\tau_{I^*,K^*}=\big[A_1(\mathbf{X}^*)\big]^\tau.
\label{eq:fullauto}
\end{equation}
Comparing the above expression, for $\tau>1$, with the observed autocorrelation function $A_\tau(\mathbf{X}^*)$ is an important test of the model.
Note that, since $-1\le A_1(\mathbf{X}^*)\le+1$, the absolute value of the autocorrelation function $\langle A_\tau\big\rangle_{I^*,K^*}$ is necessarily decreasing.
If $A_1(\mathbf{X}^*)>0$ then $\langle A_\tau\big\rangle_{I^*,K^*}$ will be positive (and exponentially decreasing) for all values of $\tau$. 
By contrast, if $A_1(\mathbf{X}^*)<0$ then $\langle A_\tau\big\rangle_{I^*,K^*}$ will be an oscillating function (modulated by a decreasing exponential), and will take negative values when $\tau$ is odd and positive values when $\tau$ is even. 

In fig. \ref{fig-5} we compare the measured autocorrelation, eq. \eqref{eq:autocorrdef}, with the predicted one, eq. \eqref{eq:fullauto}, for three different S\&P500 stocks (USB, Qcom, and MJN) over a period of 800 trading days (approximately 3.5 years). As expected, we see that the first point (one-lagged autocorrelation) is always reproduced exactly. 
We also confirm that, depending on the sign of the first point, the predicted trend is either exponentially decreasing (e.g. for the USB stock on the left) or oscillating (e.g. the Qcom and MJN stocks). 
The dashed lines indicate the noise level, that we arbitrarily fixed at two standard deviations of the Fisher-transformed \footnote{For a set of $T$ independent and identically distributed pairs of random variables $\{x_i,y_i\}_{i=1}^{T}$, the Pearson correlation coefficient $\rho_{x,y}$ is distributed around zero, but in a non-Gaussian way. However, the quantity $\phi_{x,y}\equiv \textrm{artanh}(\rho_{x,y})$, known as the \emph{Fisher tranformation}, is normally distributed around zero, with standard deviation $\sigma=(T-3)^{-1/2}$. The interval $-2\sigma<\phi_{x,y}<+2\sigma$, representing a $95\%$ confidence interval for $\phi_{x,y}$, can then be mapped back to the interval $-\tanh(2\sigma)<\rho_{x,y}<+\tanh(2\sigma)$ to obtain a $95\%$ confidence interval for $\rho_{x,y}$ around zero.} autocorrelation.
The behaviour of the USB and Qcom stocks is representative of the vast majority of stocks, with the autocorrelation within the noise level already at the minimum delay ($\tau=1$). 
This is in good agreement with what we know about financial time series (no dependencies for daily frequency, the typical time scale for autocorrelation being of the order of minutes). 
We also found that the first point, the autocorrelation between two successive days, is small but negative for most stocks in our data set. 
In the rightmost panel (MJN stock) we observe a rare dynamic, where the one-lagged autocorrelation is breaching the noise level and then rapidly oscillates to zero. 

As clear from the figure, our model reproduces well the observed autocorrelation in all these different cases, and gives a single mathematical explanation for both the exponentially decaying (from positive one-lagged autocorrelation) and the oscillating (from negative one-lagged autocorrelation) behaviour. Moreover, the generic feature of the one-dimensional Ising model, i.e. the absence of a phase transition characterized by a diverging length (here, time) scale \cite{baxter}, explains why in real-world time series the memory is always found to be short-ranged.
 
%%%%%%%%%%%%
 \begin{figure}
  \centering
  \includegraphics[width=.32\linewidth]{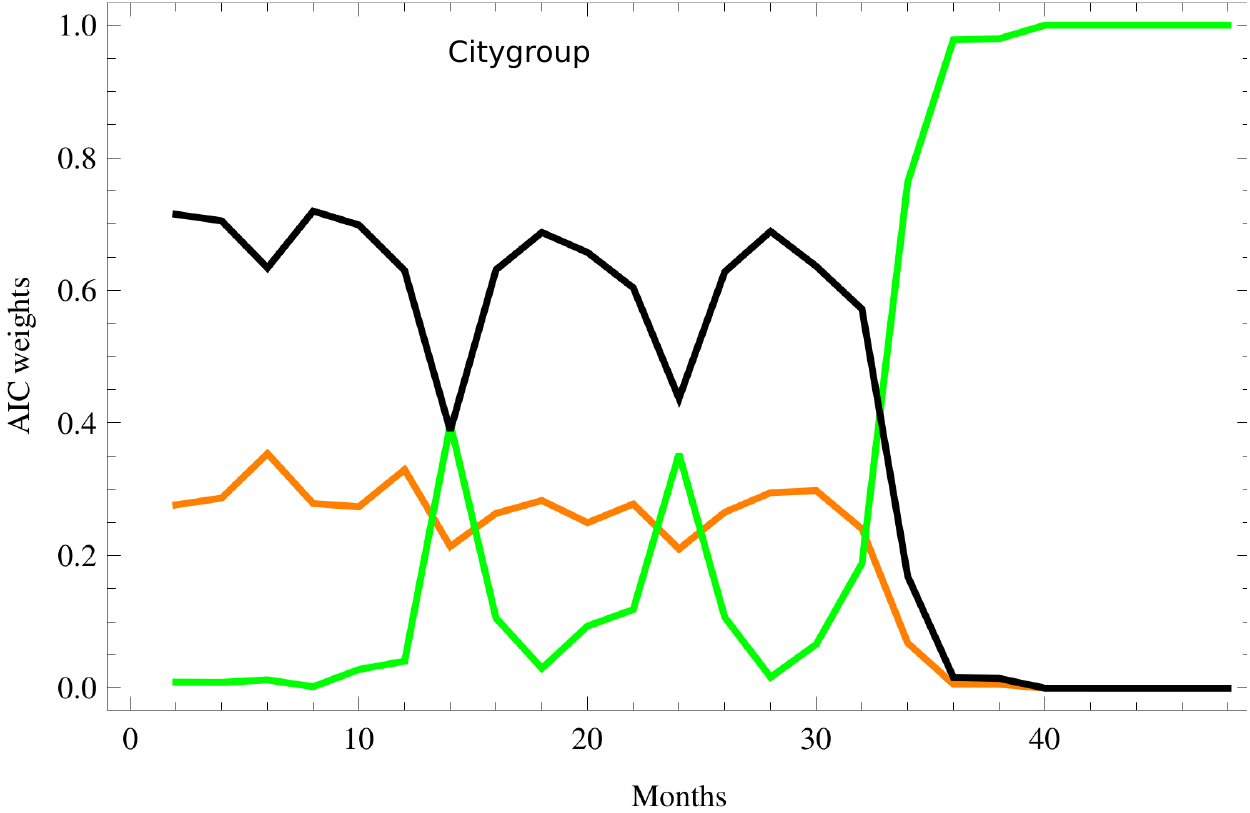}
  \includegraphics[width=.32\linewidth]{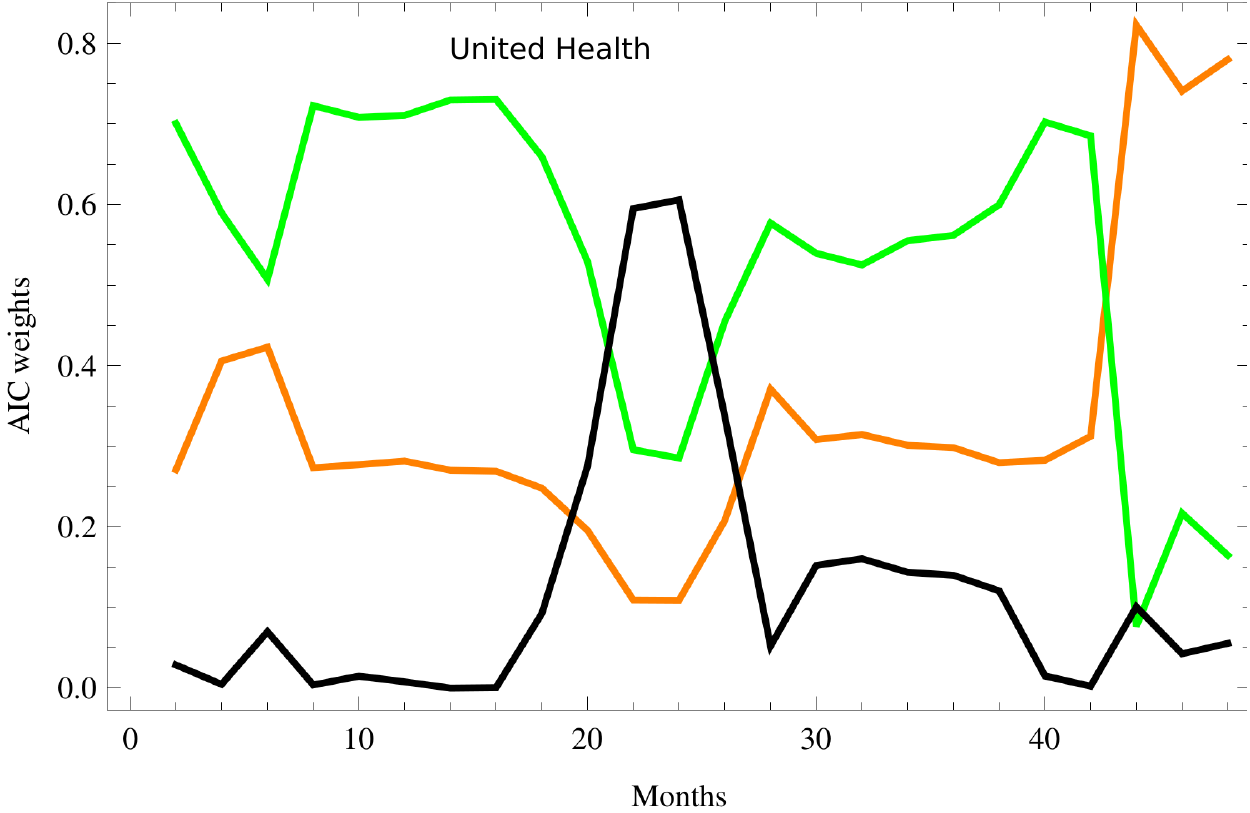}
  \includegraphics[width=.32\linewidth]{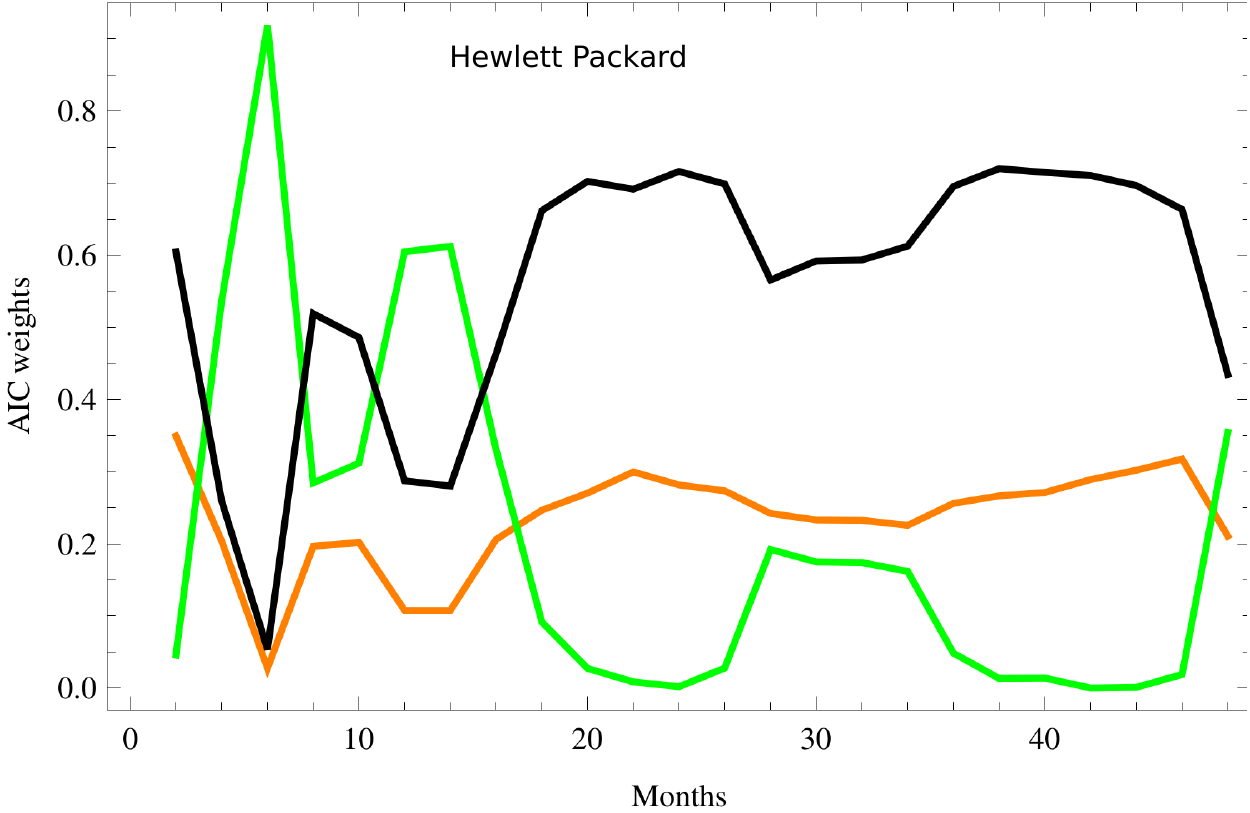}
\caption{Measured AIC weights for the three models (black: uniform random walk, orange: biased random walk, green: one-lagged model) calculated for three different S\&P500 stocks, as a function of the time horizon $T$.
The latter represents the number of months elapsed backwards from October 2011: for all stocks, all time series used to calculate the AIC weights have the same endpoint $T_0=$31 October 2011, and a variable startpoint $T_0-T$.}
\label{fig-6}
\end{figure}
%%%%%%%%%%

\subsection{Comparing the three models on empirical financial time series\label{sec:comparing}}
As we illustrated in sec. \ref{sec:AIC} in the general case, once we have more than one model for the same data $\mathbf{X}^*$, we can use the AIC weights to rank all models in terms of the achieved trade-off between accuray (good fit to the data) and parsimony (small number of parameters). The AIC weight $w_k$ of a specific model $k$ represents the probability that the model is the `best' one, among the candidate models.

We applied this procedure to the three models discussed so far (uniform random walk, biased random walk, one-lagged model).
As an example, in fig.~\ref{fig-6} we show the values of the AIC weights for three different S$\&$P500 stocks. 
We can see that the performance of the models is wildly fluctuating and different across stocks. 
This suggests that the informativeness of the measured properties is dependent on different factors, which are not entirely revealed to us. However, it is clear that in all cases the time horizon $T$ plays a key role in the performance of the models. This means that the outcome depends on how many time steps are included in the analysis.
For instance, we see that in some cases (Citigroup Inc. stock) the small $T$ regime is oscillatory, while the large $T$ regime appears to set a preference for a definite model.
In other cases (United Health Group), the three models alternate over quite long periods of time.
Most likely, this very irregular behaviour is due to the strong non-stationarity of financial markets: extending the analysis over longer time horizons does not necessarily improve the statistics, because for large $T$ the underlying price (and return) distributions change in an uncontrolled way.

We stress again that the AIC weight indicates which property, among the constraints defining all models, can better characterize the stock, given the observed data. 
In other words, it highlights the \emph{measured property} that is most informative about the original data. 
Despite the fact that the models considered so far are extremely simplified (and are by no means intended to be accurate models of financial time series), this approach can always identify, in relative terms, the most useful empirical quantity characterizing an observed time series.

%%%%%%%%%%%%%
\begin{figure*}[t]
\centerline{\includegraphics[scale=0.55]{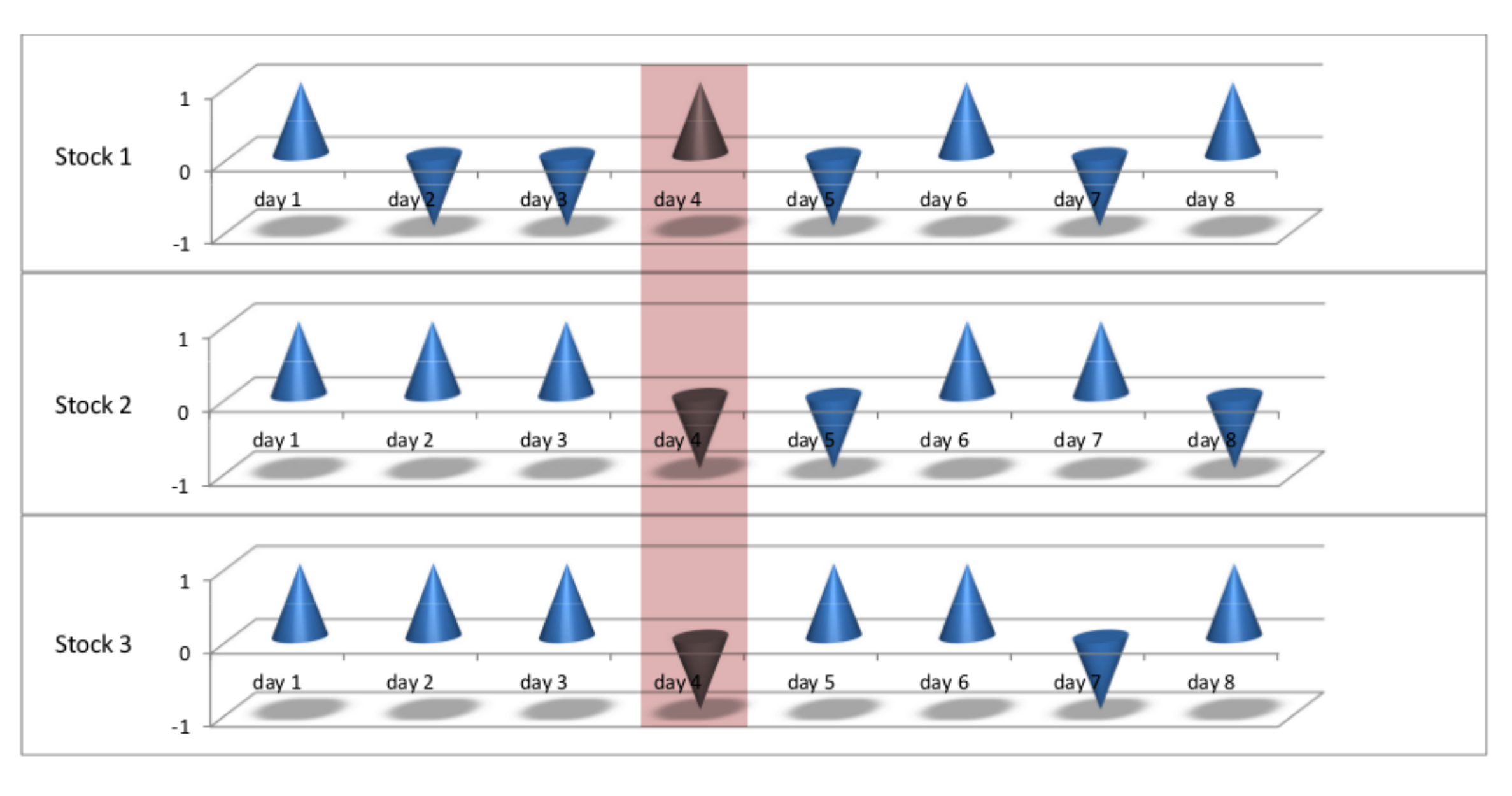}}
\caption{An example of cross section (highlighted in red) of a set of $N=3$ multiple time series. Each cross section is a $N\times1$ matrix (column vector) where each element is the instantaneous binary return of a different stock. For example, the highlighted cross section is the vector for day $t=4$.}
\label{fig:7}
\end{figure*}
%%%%%%%%%%%%

\section{Single cross-sections of multiple time series\label{sec:smts}}

In the previous section we considered models for single time series, where $N=1$ and $T$ is large. Here we consider, as a second specification of our general formalism, the somewhat `opposite' case of single cross-sections of $N$ multiple time series, which represent a daily snapshot of the market dynamics. 
For clarity, fig.~\ref{fig:7} portrays a single cross-section of a set of multiple time series.  
In this case, $T=1$ and we assume $N\gg 1$. So the matrix $\mathbf{X}$ has dimensions $N\times 1$, i.e. it is an $N$-dimensional column vector.
The entries of a cross-section $\mathbf{X}$ will be denoted by $x_i$, where $1\le i\le N$, each representing the daily increment of a different asset. 

Using again the symbol $\{\cdot\}$ to denote an average over stocks (as in sec. \ref{sec:empirical}), we now define the average increment (first moment) of $\mathbf{X}$ as
\begin{equation}
M_1(\mathbf{X})\equiv \{x_i\}=\frac{1}{N}\sum_{i=1}^{N} x_{i}
\label{eq:M1cross}
\end{equation}
and the second moment as
\begin{equation}
M_2(\mathbf{X})\equiv\{x_i^2\}=\frac{1}{N}\sum^N_{i=1} x_i^2=1.
\end{equation}
Therefore the sample variance is 
\begin{equation}
M_2(\mathbf{X})-M_1^2(\mathbf{X})=1-\{x_i\}^2.
\end{equation}
We also define the total `coupling' between stocks (for a specific cross section $\mathbf{X}$) as
\begin{equation}
D(\textbf{X})\equiv \sum_{i<j} x_{i}x_{j}=\{x_{i}x_{j}\} \frac{N(N-1)}{2}, 
\label{eq:Dcross}
\end{equation}
where now, as in eq. \eqref{eq:defcoupling}, $\{\cdot\}$ denotes an average over all pairs of stocks.

In what follows, we will consider various models for single cross-sections. The main difference with respect to the models of single time series considered in sec. \ref{sec:sts} is that the interaction between time steps for a given stock is now replaced by the interaction between different stocks for a given time step. As well known, in real financial markets the interactions among stocks (as measured e.g. via cross-correlations) are much stronger than inter-temporal autocorrelations. This makes the cross-sectional properties significantly different from those of the dynamics of single time series, once inter-stock interactions are enforced in the model. 
Yet, in simple models without interaction, we recover similar expected properties.

\subsection{Uniform random walk\label{sec:crossurw}}
As in sec. \ref{sec:singleurw}, we first consider a trivial model without constraints (see Appendix), defined by the Hamiltonian
\begin{equation}
H(\mathbf{X})=0.
\end{equation}
The probability of occurrence  of a cross section $\mathbf{X}$ is completely uniform over the ensemble of all binary cross sections of $N$ stocks. 
Again, this `gas of non-interacting spins in vacuum' model results in a uniform random walk, where all the $N$ elements of $\mathbf{X}$ are mutually independent and identically distributed. 

In the financial setting, this model assumes that all stocks fluctuate independently of each other (where the `fluctuations' are intended as ensemble ones, since we are now considering a single cross section), and under the effect of no common factor.
Each stock has zero expected value
\begin{equation}
\langle x_i\rangle=0 
\end{equation}
and maximum variance
\begin{equation}
\textrm{Var}[x_i]\equiv\langle x^2_i\rangle-{\langle x_i\rangle}^2=1.
\end{equation}

In sec. \ref{sec:comparing2}, we will compare the performance of this trivial benchmark to that of the other models we are about to introduce.
To this end, the AIC value can be calculated from eq.\eqref{eq:AIC} choosing $n_k=0$ and using the (constant) likelihood given by eq.\eqref{eq:appPcross} in the Appendix.

\subsection{Biased random walk\label{sec:crossbrw}}
In this model, which is analogous to that defined in sec. \ref{sec:singlebrw}, the constraint is chosen as the total daily increment of the cross section $\mathbf{X}$:
\begin{equation}
C(\mathbf{X})=N\cdot M_1(\mathbf{X})=N\cdot\{x_i\},
\end{equation}
where $M_1(\mathbf{X})$ is defined by eq.~\eqref{eq:M1cross}.
The Hamiltonian is then
\begin{equation}
H\left(\mathbf{X},\theta\right)=\theta \sum_{i=1}^N x_i.
\label{eq:biasbias}
\end{equation}
Similarly to its counterpart for single time series, this is a model of non-interacting spins under the effect of a common external field, and leads to a biased random walk (see Appendix).
The financial interpretation is however different: in this model, all stocks are assumed to fluctuate (again, in an `ensemble' sense) under the effect of a common market-wide factor, but are conditionally independent of each other, given the market-wide factor itself.
In the econophysics literature, the overall tendency of all stocks to move together is generally referred to as the `market mode' \cite{Econophysics book}. When applied to the data, this extremely simple model interprets the observed market mode as the consequence of an external factor (e.g. news), and not of direct interactions among stocks. 

The probability $P_i(x|\theta)$ of a given increment $x=\pm 1$ for stock $i$ is
\begin{equation}
P_i(x|\theta)= \frac{ e^{-\theta x}}{ e^{-\theta} +e^{+\theta}},
\end{equation}
the expected value of the $i$-th increment $x_i$ is 
\begin{equation}
 \langle x_i\rangle_\theta=-\tanh{\theta},
\end{equation}
and the variance is 
\begin{equation} 
\textrm{Var}[x_i] =  1- \tanh^2{\theta}.
\end{equation}

The maximum likelihood condition \eqref{eq:match}, fixing the value $\theta^*$ of the parameter $\theta$ given a real cross section $\mathbf{X}^*$, leads to 
\begin{equation} 
\theta^*=-\frac{1}{2}\ln\left[\frac{1+\{x^*_i\}}{1-\{x^*_i\}}\right],
\label{eq:fixtheta}
\end{equation}
where $\{x^*_i\}$ is the measured average increment of the observed cross section $\mathbf{X}^*$. 
We will apply this model to real financial data in secs. \ref{sec:comparing2} and \ref{sec:full}. The AIC of the model is given by eq.\eqref{eq:AIC} where $n_k=1$ and where the maximized likelihood is given by $P(\mathbf{X}^*|\theta^*)$, with $P(\mathbf{X}|\theta)$ given by eq.\eqref{eq:appPcrossbias} (see Appendix).

\subsection{Mean field model \label{sec:crossmfm}}

We now consider a more complex model, with interactions among \emph{all} stocks, which is suitable for financial cross-sections. 
Besides the constraint on the total increment, we enforce an additional constraint on the average coupling between stocks.
The resulting 2-dimensional constraint can be written as
\begin{equation}
\vec{C}(\mathbf{X})=\left(\begin{array}{c}C_1(\mathbf{X})\\C_2(\mathbf{X})\end{array}\right)=
\left(\begin{array}{c} N\cdot M_1(\textbf{X})\\ D(\textbf{X})\end{array}\right)
\end{equation}
where $M_1(\textbf{X})$ is given by eq.~\eqref{eq:M1cross} and $D(\textbf{X})$ by eq.~\eqref{eq:Dcross}.
If we write the corresponding Lagrange multiplier as
\begin{equation}
\vec{\theta}=\left(\begin{array}{c} \theta_1\\ \theta_2\end{array}\right)
=-\left(\begin{array}{c} h\\ J\end{array}\right)
\end{equation}
then the Hamiltonian reads
\begin{equation}
H(\mathbf{X},h,J)= -h\sum_{i=1}^{N} x_i - J\sum_{i<j}x_i x_j.
\end{equation}

Like the one-lagged model for single time series (see sec. \ref{sec:singleolm}), this model is formally analogous to an Ising model of interacting spins under the influence of an external `magnetic' field (here denoted by $h$). However, the big difference is that, whereas in the one-lagged model each increment $x(t)$ interacts \emph{only with the next temporal increment $x(t+1)$ of the same stock},  here each increment $x_i$ interacts \emph{with all the other increments $x_j$ of the same cross section $\mathbf{X}$}, i.e. with all other stocks in the market. 
As a model of spin systems, the above model is generally known as the mean-field Ising model \cite{baxter}. In the Appendix we provide the analytical solution of the model, adapted to our setting.

In the financial setting, this model allows us to separately consider the effects of the external field, i.e. a common factor affecting all stocks in the market, from those of the average interaction among all stocks. 
This market-wide interaction can also cause all stocks to correlate, but has the different interpretation of a collective effect, i.e. the tendency of stocks increments to `align' with each other as a result of direct interactions, rather than of a common influence. 
This is a sort of `herd effect' at the coarse-grained level of attractive ($J>0$) inter-stock interactions.
So, the model can generate the `market mode' either as the result of a common external influence such as news (in which case all stocks are still conditionally independent given the common factor), or as a collective effect due to mutual interactions (in which case all stocks are conditionally dependent given the common factor).

While the model can in principle simulate synthetic time series under a combination of the above two effects by varying the two parameters $h$ and $J$ independently, a problem arises when it is fitted to the data. 
The mathematical root of the problem is the well known fact that $H(\mathbf{X},h,J)$ can be rewritten as a linear combination of $M_1(\mathbf{X})$ and $M^2_1(\mathbf{X})$.
As we show in the Appendix, this implies that, when the maximum likelihood principle is used to fit the model to the data $\mathbf{X}^*$, the variance of $M_1(\mathbf{X})$ becomes zero. 
In other words, the model degenerates to one where $M_1(\mathbf{X})$ is no longer a random variable. This also implies that the two equations fixing the values of the parameters $J^*$ and $h^*$ become identical  (see Appendix).
Therefore it is no longer possible to uniquely fix the values of both parameters, and the problem is over-constrained. 
For this reason, we need to eliminate one parameter and consider the model only in the two extreme cases $h=0$ and $J=0$. These two cases can be treated separately. 

The case $J=0$ coincides with the biased random walk model already considered in sec. \ref{sec:crossbrw}, where $\theta=-h$. Using eq. \eqref{eq:fixtheta}, we therefore specify this model using the two parameter values
\begin{equation} 
h^*=\frac{1}{2}\ln\left[\frac{1+\{x^*_i\}}{1-\{x^*_i\}}\right],\quad J^*=0
\label{eq:spec1}
\end{equation}
where $\{x^*_i\}$ is the observed average increment of the empirical cross section $\mathbf{X}^*$. 
This model interprets the market mode as arising \emph{only} from a common external factor. 

The case $h=0$ leads us instead to a novel model where the market mode is interpreted \emph{only} as a collective effect arising from inter-stock interactions.
Using the analytical results reported in the Appendix, and in particular eq.\eqref{eq:fixJ}, we find that the parameter values are in this case
\begin{equation} 
h^*=0,\quad J^*=\frac{1}{2\{x^*_i\}(N-1)}\ln\left[\frac{1+\{x^*_i\}}{1-\{x^*_i\}}\right].
\label{eq:spec2}
\end{equation}
In what follows, when using the `mean-field' model, we will always refer to the parameter specification defined by \eqref{eq:spec2}. The other specification, eq.\eqref{eq:spec1}, will instead still be denoted as the `biased random walk' model. 

\begin{figure}[t]
\centering
\includegraphics[width=0.45\textwidth]{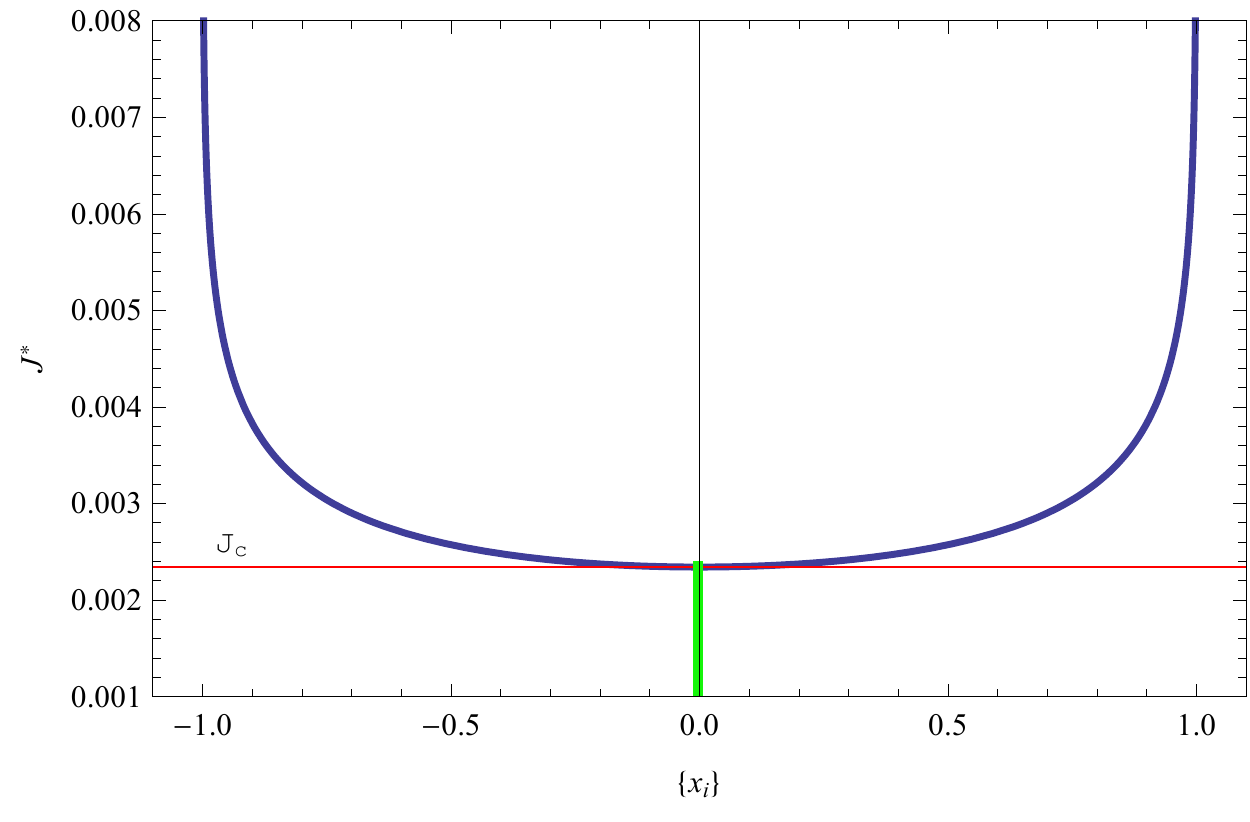}
\caption{The value of the fitted parameter $J^*$ as a function of the measured average binary return $\{x^*_i\}$ (blue curve) for a group of $N=428$ stocks (as in our S\&P sample). 
The curve shows a one-to-one relationship for $\{x_i\}\ne 0$. While $\lim_{\{x^*_i\}\to 0}J^*=J_c\equiv(N-1)^{-1}$, for $\{x^*_i\}=0$ the value of $J^*$ is actually indeterminate, as there is an infinity of values of $J^*$ (namely all values $-\infty<J^*\le J_c$, see vertical green line) that are possible solutions of the model. The value of $J_c$ is indicated by the horizontal red line.}
\label{fig-8}
\end{figure}

In fig. \ref{fig-8} we plot the value of $J^*$ as a function of $\{x^*_i\}$, as defined by eq.\eqref{eq:spec2}. We note however that eq.\eqref{eq:spec2} is undefined for $\{x^*_i\}=\pm 1$ and $\{x^*_i\}=0$.
The breakdown for $\{x^*_i\}=\pm 1$ simply means that, in order to align \emph{all} returns (in either direction), $J^*$ should diverge to $+\infty$. 
The breakdown for $\{x^*_i\}=0$ is instead more profound.
For infinitesimal (both positive and negative) values of $\{x^*_i\}$, $J^*$ admits the finite limit
\begin{equation}
\lim_{\{x^*_i\}\to 0^+}J^*=\lim_{\{x^*_i\}\to 0^-}J^*=\frac{1}{N-1}
\label{eq:limit}
\end{equation}
However, at the very point $\{x^*_i\}=0$, $J^*$ is actually indeterminate. 

%%%%%%%%%%%%%%%%%
%%%%%%%%%%%%%

The above effect is due to the well-known phase transition of the mean-field Ising model. 
In the traditional physical setting, the phase transition occurs at a critical temperature (here reabsorbed in the value of the parameters $h$ and $J$). 
When $h=0$, the critical value is obtained by setting $(N-1)J=1$, because for $(N-1)J<1$  eq. \eqref{eq:nonlinear} (see Appendix) has the single solution $\langle M_1\rangle=0$, corresponding to a phase with no macroscopic magnetization, while for $(N-1)J>1$ there are three solutions, one of which is still $\langle M_1\rangle=0$ (which is now unstable) and the other two ones being the stable solutions $\langle M_1\rangle=\pm m$ (corresponding to the onset of a macroscopic magnetization $|m|>0$ where most spins point in the same direction).
In our financial setting, since the magnetization is fixed by the data through the relation $\langle M_1\rangle=\{x^*_i\}$, the condition $(N-1)J^*=1$ implies that the phase transition occurs at the critical value 
\begin{equation}
J_c=\frac{1}{N-1}
\end{equation}
\emph{of the control parameter} $J^*$.
We can therefore rewrite eq.\eqref{eq:limit} as
\begin{equation}
\lim_{\{x^*_i\}\to 0}J^*=J_c
\end{equation}
For $J^*>J_c$ we get a `magnetized' phase where most stock prices move in the same direction (aligned returns), while for $J^*<J_c$ we get a non-magnetized phase where there is no collective alignment of stock increments, and $\{x^*_i\}=0$.
We therefore conclude that the reason why the value of $J^*$ is indeterminate for $\{x_i^*\}=0$ is because there is an infinity of values of $J^*$ (namely all values $-\infty<J^*\le J_c$) that are possible solutions of the model. 

%%% HERE
It should be noted that the case $\{x_i^*\}=0$ is never practically encountered in reality, since the empirical $\{x^*_i\}$ can be abritrarily small, but is generally not really zero. 
While this `protects' the model from the indeterminacy discussed above, it raises another problem of arbitrariness, which can however be solved very effectively using the information-theoretic criteria that we have introduced in sec.\ref{sec:AIC}. 
The problem is that the mean-field model will always interpret even the tiniest empirical deviations from $\{x_i^*\}=0$ as the result of direct interactions among stocks, and attach a value $J^*>0$ to this interpretation.
This will also apply to e.g. most realizations of a purely uniform random walk: even if for such a model one knows that the theoretical expected return is zero, most realizations will be such that $\{x_i^*\}$ is small but non-zero.
So the only phase of the mean-field model that can be explored is the `magnetized' phase dominated by collective effects.
This implies that even a pure effect of noise will be interpreted as the presence of interactions.
However, this problem will be solved in the next section, where we show that an information-theoretic comparison between the mean-field model, the uniform random walk, and the biased random walk is able to discriminate the most parsimonious model, thus allowing us to trust the mean-field model only when $\{x_i^*\}$ is distant enough from zero.

\subsection{Comparing the three models on empirical financial cross sections\label{sec:comparing2}}
We can now combine the three models together and use the AIC weights (see sec.\ref{sec:AIC}) to determine which model achieves the optimal trade-off between accuracy and parsimony.
This will immediately provide us with an indication of whether the observed market mode, as reflected in the empirical aggregate increment $\{x_i^*\}$, should be interpreted e.g. as a common exogenous factor, as a collective endogeneous effect, or even only as the sheer outcome of chance.

\begin{figure}[t]
\centering
\includegraphics[scale=0.9]{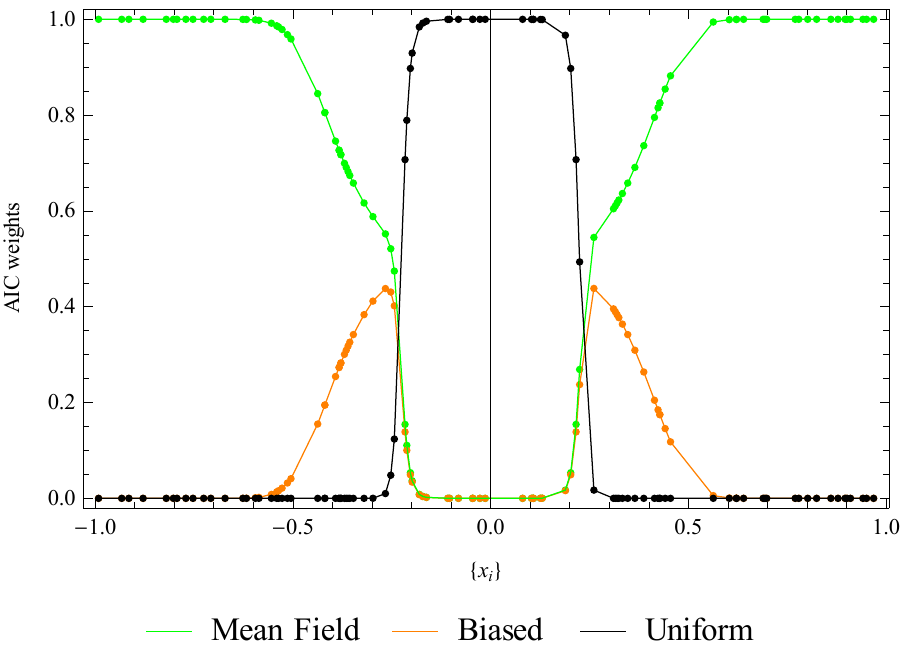}
\caption{The calculated AIC weights of the three cross-sectional models (uniform random walk, biased random walk, mean field model) as a function of the measured average daily binary return $\{x^*_i\}$, for $N=428$  S$\&$P500 stocks, each studied for 100 days of trade.}
\label{fig-9}
\end{figure}

The fact that the likelihoods of the biased random walk and the mean-field model depend only on $\{x_i^*\}$ and $N$, plus the fact that the likelihood of the uniform random walk is constant, allows us to obtain the  AIC values for the three models as functions of $\{x_i^*\}$ and $N$ only.
In fig. \ref{fig-9} we show the calculated AIC weights of the three models as a function of the observed value $ \{x^*_i\}$, for $N=428$  S\&P500 stocks.
Each point represents a different cross section, i.e. a different day of trade, for a total of 100 randomly sampled days.
It is important to note that the empirical value of the average increment only determines which point(s) of the curves are actually visited, but the curves themselves are universal.

The figure reveals us a remarkable fact, namely the presence of three distinct regimes in the behavior of the group of stocks.
For $0\leq |\{x^*_i\}| \lesssim 0.2$, we find that the best performing model is the uniform random walk, which displays an AIC weight practically equal to one (indicating that the model is almost surely the best one among the three models considered, see sec.\ref{sec:AIC}).
This means that, in this `noisy' regime, the most parsimonious explanation of the market mode, as reflected in the measured value of $\{x^*_i\}$, is that of a pure outcome of chance.

For $0.2\lesssim  |\{x^*_i\}|\lesssim 0.5$, we find that the uniform random walk is almost surely \emph{not} the best model, while the biased random walk and mean field models are competing. We observe an almost equal performance of the two models for $|\{x^*_i\}|\approx 0.2$, and an increasing preference for the mean field model as $|\{x^*_i\}|$ increases towards $0.5$. 
Despite this preference, we cannot reject the mean field model, meaning that in this `mixed' regime the most likely explanation for the market mode is a combination of exogenous and endogenous effects. 

Finally, for $0.2\lesssim  |\{x^*_i\}|\lesssim 0.5$, the mean field model achieves practically unit probability to be the best model. In this `endogenous' regime, the most likely explanation for the market model is uniquely in terms of a collective effect of direct influence among stocks.

We can summarize the above findings as follows:
\begin{equation}
\left\{\begin{array}{ll}
\textrm{Uncoordinated (noisy) regime:}&0\leq |\{x^*_i\}|\lesssim 0.2\\
\textrm{Mixed (endogenous + exogenous) regime:}& 0.2\lesssim  |\{x^*_i\}|\lesssim 0.5\\
\textrm{Coordinated (endogenous) regime:}&  0.5 \lesssim |\{x^*_i\}|\leq  1
\end{array}\right.\nonumber
\end{equation}
where we recall that the values of $|\{x^*_i\}|$ delimiting the various regimes have been calculated for $N=428$.

While the qualitative finding that larger values of $|\{x^*_i\}|$ are better explained in terms of collective effects might appear intuitive, the possibility to quantitatively identify the value $|\{x^*_i\}|\approx 0.5$ above which this intuition is fully supported by statistical evidence is a non-obvious output of the above approach. 
The same consideration applies to the identification of the other two regimes, and of a mixed phase where there is not enough statistical evidence in favour of a single interpretation of the market mode. 
Moreover, the fact that the mean field model starts being statistically significant only for $|\{x^*_i\}|\gtrsim 0.2$ solves the aforementioned problem of an otherwise problematic interpretation of even tiny values of $|\{x^*_i\}|$ as the result of inter-stock interactions.
The AIC analysis shows that, for values below $0.2$, one should not trust the mean field model, and consequently the value $J^*>0$ that the model itself indicates.
When $|\{x^*_i\}|\lesssim 0.2$, the best model is actually the uniform random walk, which effectively corresponds to $J^*=0$. This is a highly non-trivial result.

%%%%%%%%%%%%%%%
%%%%%%%%%%%%

\section{Ensembles of matrices of multiple time series\label{sec:full}}
In this section, as our third and final specification of the abstract formalism introduced in sec.\ref{sec:methods}, we extend the previous results to the general case where the observed data is a full  $N\times T$ matrix $\mathbf{X}^*$ representing a set of multiple binary time series for $N$ stocks, each extending over $T$ timesteps. 
We recall that the entries of a generic such matrix $\mathbf{X}$ are denoted by $x_i(t)$, where $i$ labels the stock and $t$ labels the time step. We assume that $N$ and $T$ are both large, i.e. $N\gg1$ and $T\gg1$.
Before introducing an explicit model, we need to make some important considerations.

We had already anticipated that the purpose of the models introduced in the previous sections was not that of introducing realistic models of financial time series. 
For instance, it is well known that the simple stochastic processes considered in sec.\ref{sec:sts} are far too simple to reproduce some key stylized facts observed in real financial time series, such as volatility clustering \cite{Sato, Krawieckia} or a bursty behavior \cite{bursty}. Moreover, being entirely binary, the above examples cannot address other well established properties characterizing the amplitude of fluctuations, e.g. the `fat' (power-law) tails of the empirical distributions of price returns.

Nonetheless, there is a simple argument that legitimates us to use a proper extension of the above modelling approach, especially that introduced in sec. \ref{sec:smts}, provided that we adequately calibrate such extension on the observed set of multiple time series. 
The argument is basically the realization that we can properly model the binary signature of a time series, using temporal iterations of even the simplistic models we have introduced in sec. \ref{sec:smts}, if we assume that some aggregated information measured on the original `weighted' time series $r_i(t)$ ($1\le i\le N$) can be used as a proxy of the driving factor defining the model itself.
We will show that this simple assumption is actually verified in the data.
In particular, we will show that a sequence of temporal iterations of the biased random walk model, which assumes that the binary time series is driven by an `external' field, can be `bootstrapped' on the real data by assuming that the field can be replaced by a function of the (endogenous) observed aggregate increment of the original weighted time series, i.e. the empirical value $\{r_i^*\}$ of the quantity $\{r_i\}$ defined in eq.\eqref{eq:defr}.
In such a way, we do not need a model generating a realistic dynamics of $\{r_i\}$ (or of the individual stock-specific increments) in order to model the behaviour of $\{x_i\}$, because the time series of $\{r_i\}$ is taken from the data. 

As a result, we will obtain an accurate model for the dynamics of the aggregate binary increment $\{x_i(t)\}$, given the observed dynamics of $\{r_i(t)\}$.
This model will reproduce with great accuracy, and mathematically characterize, the empirical non-linear relation between these two quantities that we have illustrated in sec.\ref{sec:empirical}.
We will finally test the temporal robustness and predictive power of the model, and conclude with discussion of the relatedness of our approach and more traditional `factor models' in finance.

%%%%%%%%%%%%%%%%%
\begin{figure*}[t]
\centerline{\includegraphics[width=.99\linewidth]{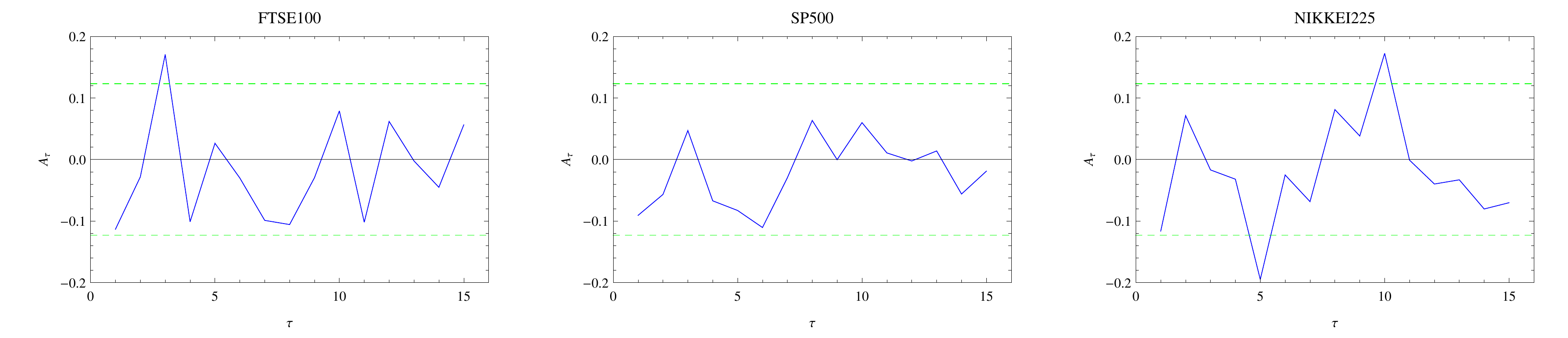}}
\caption{The measured autocorrelation of the average binary daily return $\{x^*_i(t)\}$ for the three indices in year 2006.  The green lines represent the noise level, calculated as $\pm 2$ standard deviations of the Fisher-transformed autocorrelation.}
\label{fig-10}
\end{figure*}
%%%%%%%%

\subsection{Temporal dependencies among cross sections}
In order to execute the above plan, we first analyze the correlations between single cross sections of the market. 
We need this preliminary analysis in order to determine whether the temporal extension of the models defined in sec. \ref{sec:smts} should incorporate dependencies among different snapshots.

Based on extensive financial literature, we expect no correlation (on a daily frequency) among the returns of different cross sections. 
However, most analyses focus on the auto-correlation of \emph{individual} stocks, based on their \emph{weighted} returns.
So, to check our hypothesis we perform an explicit analysis of the temporal auto-correlation of the observed time series of the \emph{aggregate}, \emph{binary} return $\{x^*_i(t)\}$. 
This analysis is shown in fig.\ref{fig-10} for the three indices, using daily data for year 2006. 
We confirm that the observed autocorrelation is not statistically significant, since (apart for a few points) it lies within the range of random noise (calculated by imposing a threshold of two standard deviations on the Fisher-transformed autocorrelation).
This type of uncorrelated dynamics is observed throughout our dataset.
This means that, in line with other analyses of autocorrelation, the memory of the aggregate binary return of real markets, if any, is much shorter than a day. 

Going back to the result illustrated in fig. \ref{fig-9}, we can then conclude that there is no significant correlation in the trajectories of the daily points populating the curves. 
In other words, given the knowledge of the position of the market in the AIC curves in a given day, we cannot predict where the market will move the next day, even if of course we know that it will move to another point in the curves themselves.

\subsection{Reproducing the observed binary/non-binary relationships\label{sec:reproducing}}
The previous result sets the stage for our next step, where we consider an explicit extension of the models considered in sec.\ref{sec:smts} to an ensemble of multiple time series, as introduced in sec.\ref{sec:methods} in the general case. 
The absence of autocorrelation implies that we can define the Hamiltonian of the full $N\times T$ matrix $\mathbf{X}$ as a sum of $T$ non-interacting Hamiltonians, each describing a single cross section of $N$ stocks.

Next, we need to choose the model to extend. We want the final model to establish (among other things) an expected relationship between the binary and the weighted aggregate returns, so that we can test this prediction against the empirical relationships illustrated in sec.\ref{sec:empirical}. 
This implies that we need to input the measured weighted return $\{r^*_i\}$ as a driving parameter of the binary model. 
Among the three models, only the biased random walk and the mean field model have parameters that can be related to $\{r^*_i\}$.
In sec. \ref{sec:smts} we treated those models as giving competing interpretations of the market model in terms of exogenous and endogenous effects respectively. However, it should be noted that this is no longer possible as soon as the parameters of these models are made dependent on the observed return.
For instance, if we assume that the parameter $\theta$ of the biased random walk depends on $\{r^*_i\}$ (which is a property of the data), we can no longer interpret $\theta$ as an external field, since it has been somehow `endogenized'. 
Determining whether $\theta$ can be interpreted as endogenous or exogeneous is now entirely dependent on whether $\{r^*_i\}$ itself can be interpreted as endogenous or exogeneous. 
This tautology does not prevent us from determining a relationship between $\{r^*_i\}$ and $\{x^*_i\}$ in their full range of variation, because such relationship is independent on the optimal (endogenous or exogenous) interpretation of both quantities. 

We also note that the choice of the model to calibrate on $\{r_i\}$ is now completely independent of the relative performance of the various models that we have determined in the case of free parameters, including their AIC weights shown in fig.\ref{fig-9}. 
Indeed, apart from an initial calibration, the parameters will no longer be fitted using the maximum likelihood principle, making the AIC analysis no longer appropriate. 
In other words, ranking the `free' models and endogenizing their parameters are two completely different problems. In particular, the low AIC weight of the biased random walk throughout most of fig.\ref{fig-9} does not impede us from using this model in our next analysis.
We will indeed `bootstrap' the biased random walk on the real data, by looking for a relationship between $\{r_i\}$ and the parameter $\theta$. We prefer this model over the mean field one because, while it is natural to think of (a function of) $\{r_i\}$ as a proxy of the `field' $\theta$ affecting the market in the biased random walk model (notably, $\{r_i\}$ has a definition similar to that of a market index), it is less natural to think of the same quantity as a proxy of the inter-stock interaction $J$ in the mean field model (although, as we said before, this would be technically possible).
 
Combining all the above considerations, we finally generalize the biased random walk model defined by eq.\eqref{eq:biasbias} to the matrix case as follows:
\begin{equation}
H(\mathbf{X},\vec{\theta})=\sum_{t=1}^{T} \theta(t)\sum_{i=1}^{N}x_i(t)
\label{eq:superH}
\end{equation}
where $\vec{\theta}$ it a $T$-dimensional vector with entries  $\theta(t)$. 
Note that, while the models we introduced in sec.\ref{sec:sts} have time-independent parameters and therefore correspond to time series at statistical equilibrium (for example a model with constant volatility), we are now considering more general models with time-dependent parameters. 
Relating $\theta(t)$ to $\{r_i(t)\}$ will allow us to incorporate any observed degree of non-stationarity of the data into the model itself.

As a preliminary calibration, we now look for an empirical relation between $\{r_i(t)\}$ and $\theta(t)$. To this end, we first treat the latter as a free parameter and look for the optimal value $\theta^*(t)$ maximizing the likelihood of the observed binary time series $\mathbf{X}^*$.
Since the Hamiltonians for different timesteps are non-interacting, it is easy to show that $\theta^*(t)$ is given again by eq.\eqref{eq:fixtheta} where $\{x_i^*\}$ is replaced by $\{x_i^*(t)\}$:
\begin{equation} 
\theta^*(t)=-\frac{1}{2}\ln\left[\frac{1+\{x^*_i(t)\}}{1-\{x^*_i(t)\}}\right].
\label{eq:fixthetat}
\end{equation}

%%%%%%%%
\begin{figure*}[t]
\centerline{\includegraphics[width=.99\linewidth]{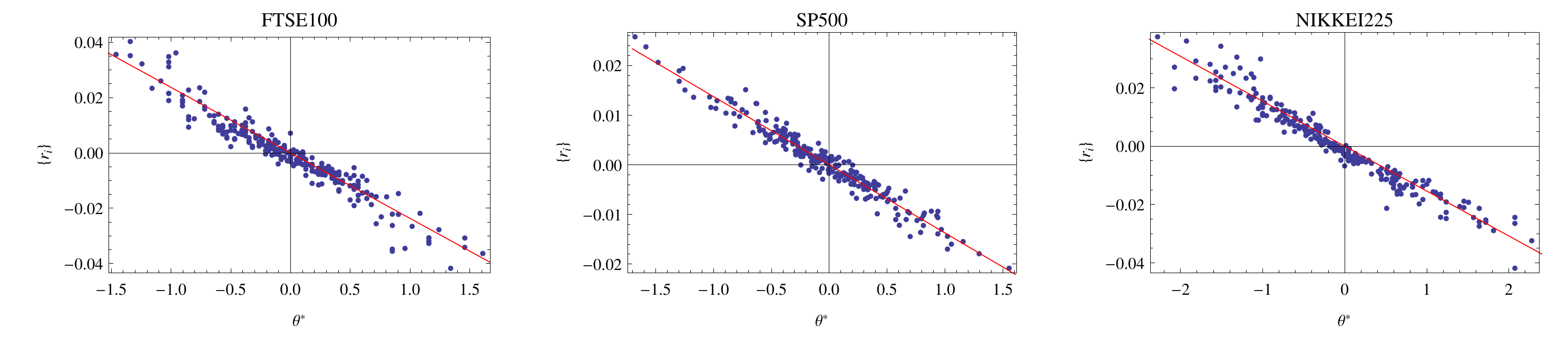}}
\caption{The most likely value of the driving field $\theta^ *(t)$ calculated applying the biased random walk model to the projected binary signature of day $t$, compared with the measured average weighted return $\{r^*_i(t)\}$ of the same day, for 250 trading days (approximately one year) in the FTSE100 (left), S\&P500 (center) and NIKKEI225 (right) in various years (2003, 2007, and 2004 respectively). We also show the linear fit $\{r_i(t)\}=-c~\theta^*(t)$ with $c>0$.}
\label{fig-11}
\end{figure*}
%%%%%%%%%%%%

In fig. \ref{fig-11} we compare the resulting value of $\theta^*(t)$ with the corresponding observed weighted return $\{r^*_i(t)\}$, for the three indices separately. 
Each point in the plot corresponds to a different day, and we considered 250 days (approximately one year) for each index. We find a strong linear relation between the two quantities. This relation can be fitted by the one-parameter curve 
\begin{equation}
\{r^*_i(t)\}=- c\cdot\theta^*(t)
\label{eq:fitc}
\end{equation}
where $c>0$. 
This finding is very important. 
It confirms that the parameter $\theta^*(t)$, defined through eq.\eqref{eq:superH} as a time-varying `field' driving the observed binary increment $\{x^*_i(t)\}$ with maximum likelihood, is an excellent proxy for the observed non-binary `market index' $\{r^*_i(t)\}$. This result holds up to a negative factor $c$ which, on the time scale considered, is constant for each market (in sec. \ref{sec:c} we will provide a more detailed analysis of the stability of $c$ over different time scales). 
Since $\{r^*_i(t)\}$ is a property measured on the stock increments themselves, it reflects both external influences and internal dependencies. Therefore $\theta^*(t)$ cannot be (entirely) interpreted as an external field. This confirms our interpretation of the biased random walk as a model agnostic to the (endogenous or exogenous) nature of the driving field in the present setting.

\begin{figure*}[t]
\centerline{\includegraphics[width=.99\linewidth]{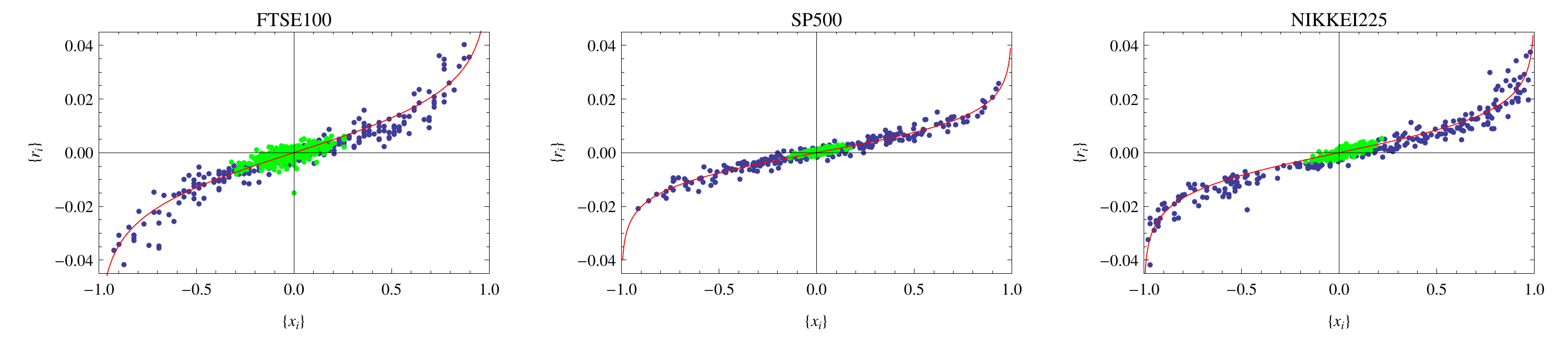}}
\caption{Nonlinear relationship between the average daily increment (weighted return) and the average daily sign (binary return) over all stocks in the FTSE100 (left), S\&P500 (center) and NIKKEI225 (right) in various years (2003, 2007, and 2004 respectively). 
Here each point corresponds to one day in the time interval of 250 trading days (approximately one year). The red curve is our non-parametric prediction based on the fit shown in fig.\ref{fig-11}, and the green points are the same properties measured on the shuffled data.}
\label{fig-12}
\end{figure*}

Combining eqs.\eqref{eq:fixthetat} and \eqref{eq:fitc} together, we finally obtain a mathematical expression for the expected relationship between $\{r^*_i\}$ and $\{x^*_i\}$ in our model:
\begin{equation} 
\{r^*_i(t)\}=\frac{c}{2}\ln\left[\frac{1+\{x^*_i(t)\}}{1-\{x^*_i(t)\}}\right]=c\cdot\textrm{artanh}\{x^*_i(t)\}.
\label{eq:endo}
\end{equation}
Inverting, we have 
\begin{equation} 
\{x^*_i(t)\}=\tanh\frac{\{r^*_i(t)\}}{c}.
\label{eq:endo2}
\end{equation}
We can now test the above expressions against the data shown previously in fig.\ref{fig:empir1}. In that figure, we already showed that the observed relationship between $\{r^*_i\}$ and $\{x^*_i\}$ can be fitted very well by a curve of the form given by eq.\eqref{eq:endo}. 
We have just provided a theoretical justification for the otherwise arbitrary use of such expression. 
Moreover, now we can fit the value of $c$ using eq.\eqref{eq:fitc}, which is independent of eq.\eqref{eq:endo}. 
Once we obtain $c$ in this way, we can use eq.\eqref{eq:endo} to predict $\{r^*_i(t)\}$ given $\{x^*_i(t)\}$, or \emph{vice versa}, without fitting any parameter. 
In fig.\ref{fig-12} we show the result of this operation. 
We confirm that the prediction of our model matches the empirical relationship very well. 

We also consider a null model where we randomly shuffle the increments of each of the $N$ time series independently. This results in a set of randomized time series, with elements $r'_i(t)$, where the total increment $\sum_{t=1}^Tr'_i(t)$ for each stock is preserved, but the returns of all stocks in a given day are uncorrelated.
From $r'_i(t)$, we obtain the binary signature $x'_i(t)$ as for the real data.
As shown in fig.\ref{fig-12}, this randomized benchmark overlaps with the empirical trend only in a very narrow, linear regime.
We will now try to understand this result.

The reason why the shuffled data result in a linear trend is the following. 
For each value of $\{x'_i\}$, there is a definite number $N_{up}$ of `up' stocks and a definite number $N_{down}=N-N_{up}$ of `down' stocks, according to the relation
\begin{equation}
\{x'_i\}=\frac{N_{up}-N_{down}}{N}=\frac{2N_{up}-N}{N}=2\frac{N_{up}}{N}-1.
\end{equation}
Conditional on the above value of $\{x'_i\}$, the expected value of $\{r'_i\}$ (over multiple shufflings) is
\begin{equation}
\langle\{r'_i\}\rangle=\frac{r^*_+N_{up}+r^*_-N_{down}}{N}\approx\frac{r^*_+N_{up} -r^*_+N_{down}}{N}=r^*_+\left[2\frac{N_{up}}{N}-1\right]=r^*_+\{x'_i\}.
\label{eq:linear}
\end{equation}
where $r^*_+>0$ is the average positive increment (over all $T$ time steps and all $N$ time series) and $r^*_-<0$ is the average negative increment. Note that both values coincide with the corresponding quantities in the original data, and have been denoted by a star accordingly.
Assuming approximately symmetric log-return distributions for each of the $N$ time series as typically observed, we have set $r^*_-\approx- r^*_+$. Given the overlap between real and shuffled data around zero returns in fig.\ref{fig-12}, we can linearize eq. \eqref{eq:endo2} around zero and compare it with eq.\eqref{eq:linear} to get 
\begin{equation}
c\approx r^*_+.
\end{equation}
The above expression suggests that the value of $c$ strongly depends on the original log-return distribution. Therefore, we expect that the stability of $c$ is determined by that of $r^*_+$. 
In sec. \ref{sec:c} we will study the stability of $c$ in more detail.

The above simple argument shows that, for shuffled data, we indeed expect a linear relationship between $\{r'_i(t)\}$ and $\{x'_i(t)\}$.
This is a striking difference with respect to real data, where $\{r^*_i(t)\}$  virtually diverges as $|\{x^*_i(t)\}|$ approaches one.
This `divergence' indicates that, when most stocks are aligned in real markets ($|\{x^*_i(t)\}|\approx 1$), the observed log-returns are much larger than the typical positive increment ($|\{r^*_i(t)\}|\gg r^*_+$). In other words, extreme log-returns are more often observed when stocks are synchronized.
This means that there is a strong correlation between the magnitude of log-returns of individual time series and the degree of coordination of all stocks in the market.

While for infinite realizations of the shuffling procedure we would observe eq.\eqref{eq:linear} extending to the full range $-1\le \{x' _i\}\le +1$, for finite realizations we observe a much narrower span of values (see fig.\ref{fig-12}). 
This is due to the absence of correlations among stocks, resulting in significantly lower values of both $\{r'_i\}$ and $\{x'_i\}$ with respect to the observed quantities $\{r^*_i\}$ and $\{x^*_i\}$.
Interestingly enough, for the S\&P500 index the randomized data span the range $|\{x'_i\}|\lesssim 0.2$, which coincides precisely with the regime we identified in fig.\ref{fig-9} for a completely noisy-driven system with the same number of stocks.
This confirms that the AIC analysis correctly pinpoints the boundaries outside which one should expect the observed value $\{x^*_i\}$ to be inconsistent with a typical realization of $N$ purely random variables.

%%%%%%%%%%%%%%%%%
%%%%%%%%

\begin{figure*}[t]
\centerline{\includegraphics[width=.99\linewidth]{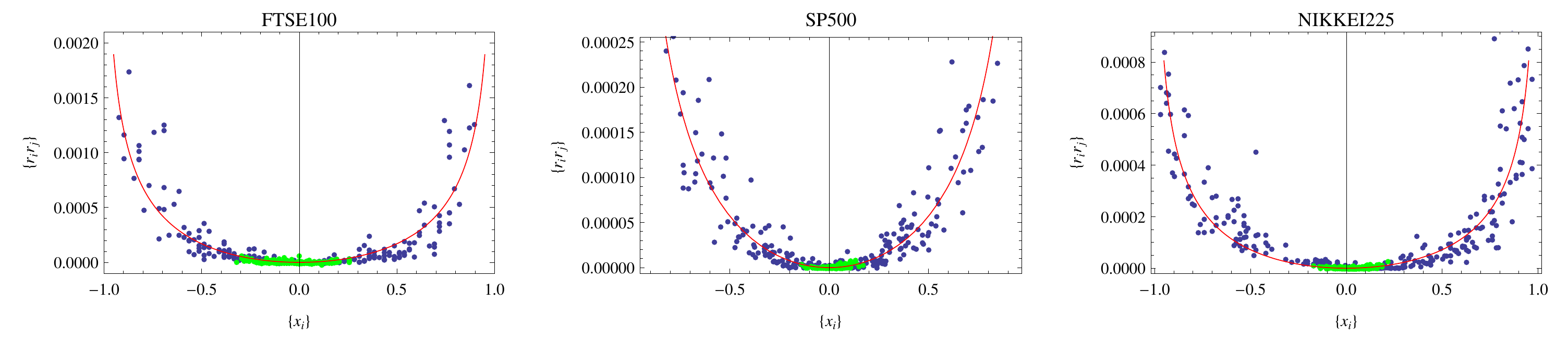}}
\caption{Nonlinear relationship between the average daily coupling (weighted coupling) and the average daily sign (binary return) over all stocks in the FTSE100 (left), S\&P500 (center) and NIKKEI225 (right) in various years (2003, 2007, and 2004 respectively). 
Here each point corresponds to one day in the time interval of 250 trading days (approximately one year). The red curve is our non-parametric prediction based on the fit shown in fig.\ref{fig-11}, and the green points are the same properties measured on the shuffled data.}
\label{fig-13}
\end{figure*}

The above results also provide an explanation for the second empirical nonlinear relation that we had documented in sec. \ref{sec:empirical}, i.e. the one between $\{r^*_i(t)r^*_j(t)\}$ and $\{x^*_i(t)\}$ (see fig. \ref{fig:empir2}).
In general, we can write $\{r_ir_j\}$ as
\begin{equation}
\{r_ir_j\}=\frac{1}{N(N-1)}\sum_{i\ne j}r_i r_j=\frac{1}{N(N-1)}\left[\left(\sum_{i=1}^Nr_i\right)^2-\sum_{i=1}^Nr_i^2\right].
\end{equation}
The term $\sum_{i=1}^Nr_i^2$ is of order $N$, and vanishes for large markets when divided by $N(N-1)$. We are therefore left with
\begin{equation}
\{r_ir_j\}\approx\frac{1}{N(N-1)}\left(\sum_{i=1}^Nr_i\right)^2\approx \{r_i\}^2.
\end{equation}
Using eq.\eqref{eq:endo}, we get
\begin{equation}
\{r^*_i(t)r^*_j(t)\}\approx\{r^*_i(t)\}^2=c^2\textrm{artanh}^2\{x^*_i(t)\}
\end{equation}
which theoretically justifies the fitting function we had used in fig.\ref{fig:empir2}.
Again, rather than fitting that curve on the data, we can use the value of $c$ determined from the (independent) fit shown in fig.\ref{fig-11}. 
This results in the non-parametric plot shown in fig. \ref{fig-13}.
We confirm that, for each of the three indices, we can reproduce the observed relationship very well.

As before, we also show the relationship between $\{r'_i(t)r'_j(t)\}$ and $\{x'_i(t)\}$ for randomly shuffled data.
The linearity of eq.\eqref{eq:linear} now translates into an expected parabolic relationship:
\begin{equation}
\langle\{r'_ir'_j\}\rangle\approx\{r'_i\}^2=(r^*_+)^2\{x'_i\}^2.
\label{eq:parabola}
\end{equation}
Again, real data strongly deviate from the above `uncorrelated' parabolic expectation, because extreme events make the empirical coupling $\{r^*_ir^*_j\}$ virtually `diverge' when stocks are highly synchronized ($|\{x^*_i\}|\approx 1$).

%%%%%%%%
%%%%%%%%%%%

\subsection{Stability of the parameter $c$\label{sec:c}}

Once we have mathematically characterized the observed nonlinear relations, an unavoidable question arises: in a given market, how stable are those relations? 
Since $c$ is the only parameter in the above analysis, the question simply translates into the stability of $c$.
We have already noted that $c$ is related to the average positive return $r_+^*$, which we expect to be relatively stable.
In order to study the stability of $c$ in more detail, we now consider several yearly and monthly time windows, and explore the time evolution of the fitted parameter for the three indices.

\begin{figure*}[t]
\centerline{\includegraphics[width=.99\linewidth]{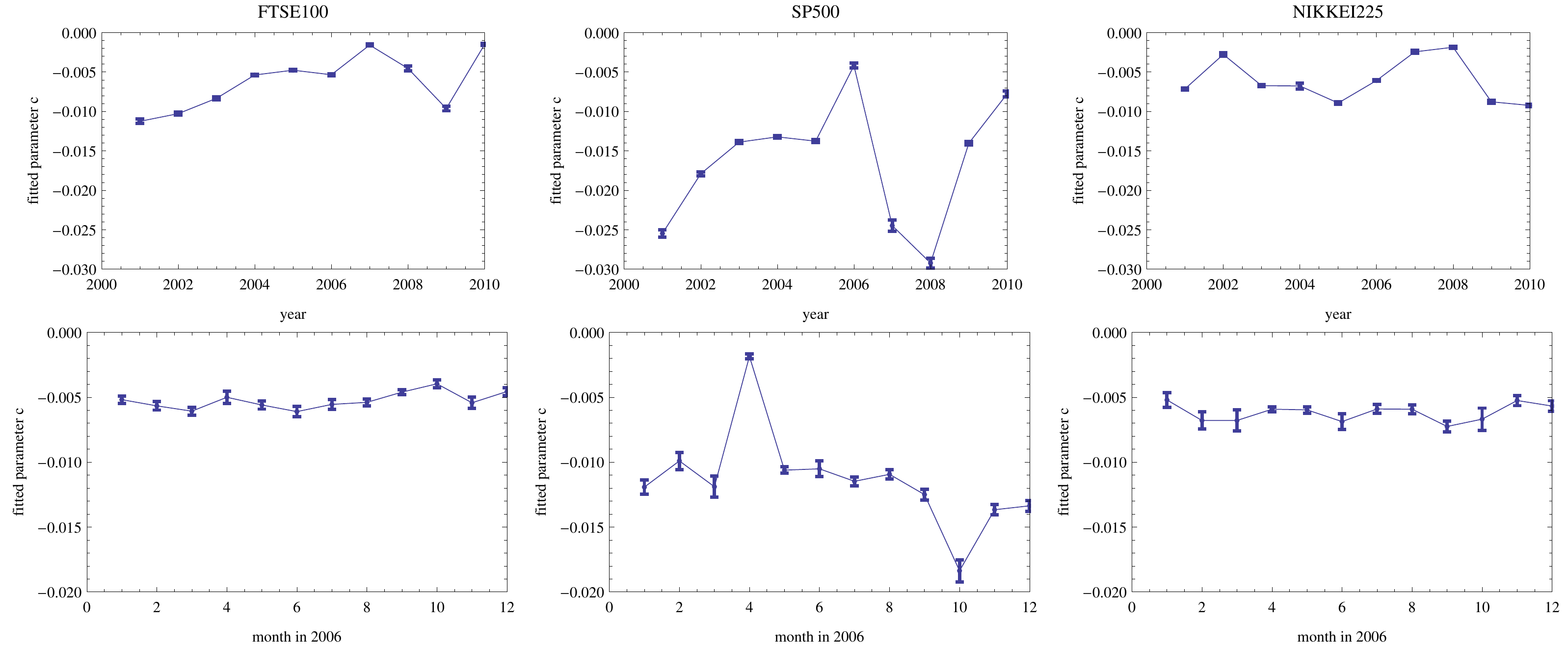}}
\caption{Stability of the parameter $c$, fitted as in fig.\ref{fig-11} on various yearly (top panels) and monthly (bottom panels) snapshots of the market, for the FTSE100 (left), S\&P500 (center) and NIKKEI225 (right).}
\label{fig-14}
\end{figure*}

In fig. \ref{fig-14} (upper panels) we plot the values of the parameter $c$ (with error bars) for 11 yearly snapshots (2001-2010).
It is clear that there are periods during which the yearly values are relatively stable, and periods when they fluctuate wildly. Thus, in most cases the fitted value of $c$ in a given year does not allow to make predictions about the value of $c$ int the next year.

However, we can also consider a monthly frequency. In the bottom panels of fig. \ref{fig-14} we show the result of our analysis, when carried out on the 12 monthly snapshots of year 2006. 
We choose this particular year because, in the yearly trends shown above, it represents very different points for different markets: the end of a stable period for the FTSE100, an exceptional jump for the S\&P500, and the middle of an increasing trend for the NIKKEI225.
Despite these differences, we find that in all three markets the monthly dynamics is much more stable than the yearly one.
In particular, the trends for FTSE100 and NIKKEI225 are almost constant, and for the S\&P500 there are only two deviating points from an otherwise stable trend (despite the large fluctuation that 2006 represents in the yearly trend for this index). 
This implies that, in most cases, one might even use the monthly value of $c$ out of sample, in order to predict the future relationship between $\{x_i\}$ and $\{r_i\}$ based on a past observation.
We should however stress that the aim of our method is to characterize such relationship, and not to predict it.
Indeed, we cannot imagine any situation in which only the binary (or only the non-binary) information is available.

The above results show that there is a trade-off between short and long periods of time. For short (e.g. monthly) periods there are less points to calculate $c$ through a fit of the type shown in fig.\ref{fig-11}. This explains why the monthly trends in fig.\ref{fig-14} have bigger error bars than the yearly trends in the same figure. 
By contrast, for longer (e.g. yearly) periods each individual fit is better, but there are more fluctuations in the temporal evolution of the parameter $c$, because the data are less stationary. 
In general, we expect that in each market, and for a specific period of time, there is a different `optimal' frequency to consider.

 %%%%%%%%
%%%%%%%%%%%

\subsection{Relation to factor models}

We would like to conclude this paper with a discussion of the relationship between some of our findings and the popular \emph{factor models} in the financial literature \cite{Bouchaud1}. As a basic consideration, we stress that factor models can only be applied to the original (non-binary) increments (it is impossible to decompose a binary signal into a nontrivial combination of binary signals), while our models only apply to the binary projections. We should bear this irreducible difference in mind in what follows. However, due to the mapping between binary and non-binary increments that we have documented, we can try to indeed relate the two approaches.

First, let us consider the shuffled (uncorrelated) data, where the original log-returns are randomly permuted within each of the $N$ time series. 
It is well known that the total temporal increment (over $T$ time steps) of any empirical time series of price increments is generally close to zero (due to market efficiency), and that the distribution of log-returns is mostly symmetric around this value. 
This is especially true if each of the $N$ original time series has been separately standardized, i.e. the $i$-th temporal average has been subtracted from each increment of the $i$-th time series, and the result has been divided by the $i$-th standard deviation. 
In such a case, the $N$ log-return distributions become also very similar to each other, because their support is the same and their values are comparable. 
This means that, after the shuffling, the time series are sequences of independent and almost identically distributed variables with zero mean. 
We denote the corresponding increments as 
\begin{equation}
r_i(t)= \epsilon_i(t)\quad \forall i,
\label{eq:zerofactor}
\end{equation} 
where the $\epsilon_i$'s are random variables. In a traditional factor analysis, the above scenario takes the form of a `zero-factor' model. 
Under this model, the aggregate increment over $N$ stocks is expected to be narrowly distributed around
\begin{equation}
\{r_i(t)\}= \frac{1}{N}\sum_{i=1} ^N \epsilon_i(t)\approx 0.
\end{equation}
When $\{r_i(t)\}$ takes small values around zero, we know from fig. \ref{fig-12} that $\{x_i(t)\}$ also takes small values around zero. 
Indeed, shuffled time series are in the linear regime that spans the range where the binary increment $\{x_i(t)\}$ is consistent with a uniform random walk (see fig.\ref{fig-9}).
Therefore we find that the zero-factor model (for the non-binary returns) and the uniform random walk (for the binary returns) are consistent with each other in the linear regime. In other words, when in our analysis we measure a value of $\{x_i(t)\}$ that is consistent with a uniform random walk, we know that the original log-returns are consistent with a zero-factor model.

Next, we consider a one-factor model, where there is one dominant underlying factor assumed to control the dynamics of all the time series.
In such a case, each return can be decomposed as
\begin{equation}
r_i(t)=\alpha_i \Phi_0(t)+\epsilon_i(t)\quad \forall i,
\label{eq:onefactor}
\end{equation} 
where $\alpha_i$ is the `factor loading' of the $i$-th time series with the  dominant factor $\Phi_0(t)$. 
When referring to stocks, the factor $\Phi_0(t)$ is attributed to the market mode. 
It is known that, during crisis times when the markets are highly correlated, a one-factor model can describe the dynamics quite well.
Under this model, the aggregate increment is
\begin{equation}
\{ r_i(t)\}=\frac{1}{N} \sum_{i=1} ^N \alpha_i \Phi_0(t)+\frac{1}{N}\sum_{i=1} ^N\epsilon_i(t) \approx \{ \alpha_i \} \Phi_0(t)
\end{equation} 
where $\{ \alpha_i \}\equiv\frac{1}{N}\sum_{i=1} ^N \alpha_i$ is the average loading, which is independent of both $i$ and $t$. 
This result implies that, when the market is well described by a one-factor model, the average increment $\{ r_i(t)\}$ that we measure in our analysis is proportional to the factor $\Phi_0(t)$ itself. 
We note that the one-factor model is somehow similar to our biased random walk model, as it assumes a common drive for all the stocks.
However, since $\Phi_0(t)$ is fitted on the data, the one-factor model cannot distinguish between an endogenous or exogenous nature of the common drive.
This situation is similar to when we use the observed value of $\{ r_i(t)\}$ as the driving field of the biased random walk (see sec.\ref{sec:reproducing}). 

In financial analysis, the factor model can be used to filter the original time series and remove the one-factor component from them. 
When the model is a good approximation to the real market, the filtered returns are $r_i(t) \approx \epsilon_i(t)$, leading us back to eq.\eqref{eq:zerofactor} and the related considerations. 
In such a scenario, there is no correlation among the stocks, and each stock is acting as an i.i.d. variable.
We therefore expect that, if we remove the market mode from the original time series, then (in periods where the market is indeed dominated by a single factor) we would obtain results similar to the shuffled case, and we would find the system in the uncoordinated phase of fig.\ref{fig-9}.

However, despite the fact that in certain conditions the one-factor model can generate the market behaviour, the model is too simplistic \cite{Bouchaud1}. 
In reality the dynamics is more complex and can be attributed to many factors, that sometimes overlap with industrial (sub)sectors.
Generally the different factors are identified by the largest, non-random eigenvalues of the empirical cross-correlation matrix, where the market mode relates to the highest eigenvalue \cite{Bouchaud1}.
The presence of many deviating eigenvalues is an indication of the fact that the one-factor model should be rejected.
A more realistic, $M$-factor model is
\begin{equation}
r_i(t)=\sum_{j=0}^M\alpha_{ij} \Phi_j(t)+\epsilon_i(t)\quad \forall i,
\label{eq:multifactor}
\end{equation} 
where $j=0$ denotes a common market-wide factor as above, while $j>0$ denotes sector-specific factors.
In such a case, our measured value of $\{ r_i(t)\}$ is
\begin{equation}
\{ r_i(t)\}=\frac{1}{N} \sum_{i=1} ^N   \sum_{j=1}^M \alpha_{ij} \Phi_j(t)+\frac {1}{N}\sum_{i=1} ^N\epsilon_i(t)\approx \sum_{j=1}^M\{ \alpha_{ij} \} \Phi_j(t)                             
\end{equation} 
which is a linear combination of the multiple factors controlling the market dynamics.

It should be noted that factor models cannot distinguish between an endogenous and exogenous origin for the factors $\Phi_j(t)$ themselves, even if we invoke some information-theoretic criterion to rank different specifications of these models. 
By contrast, our binary models allow us to discriminate among these multiple scenarios, as we have shown in fig.\ref{fig-9} and related discussions.
Moreover, while our approach allows us to relate binary and non-binary increments of real time series and replicate the observed relationships among them (see figs.\ref{fig-12} and \ref{fig-13}), factor models cannot lead to a similar result, because they do not allow for a binary description.

\section{Conclusions\label{sec:Conclusions}}

We presented a novel method for the analysis of single and multiple binary time series. Our information-theoretic approach allowed us to extract and quantify the amount of information encoded in simple, empirically measured properties. This resulted in the possibility to associate an entropy value to a time series given its measured  properties, and to compare the informativeness of different measured properties.

By employing our formalism, we have identified distinct regimes in the collective behavior of groups of stocks, corresponding to different levels of coordination that only depend on the average return of the binary time series. In each regime the market exhibits a dominant character: the market mode can be interpreted as an exogenous factor, as pure noise, or as a combination of endogenous and exogenous components. Moreover, each regime is characterized by the most informative property.

Finally and more importantly, we were able to replicate the observed non-linear relations between binary and non-binary aggregate increments of real multiple time series.
We have mathematically characterized these relations accurately, and interpreted them as the result of the fact that very large log-returns occur more often when most stocks are synchronized, i.e. when their increments have a common sign.
Our findings suggest that the binary signatures carry significant information, and even allow to measure the level of coordination in a way that is unaccessible to standard non-binary analyses.

\begin{acknowledgments}
We thank Marc van Kralingen for a thorough reading of our manuscript and for identifying some mistakes.
We acknowledge support from the Dutch Econophysics Foundation (Stichting Econophysics, Leiden, the Netherlands) with funds from beneficiaries of Duyfken Trading Knowledge BV, Amsterdam, the Netherlands.
This work was also supported by the EU project MULTIPLEX (contract 317532) and the Netherlands Organization for Scientific Research (NWO/OCW).
\end{acknowledgments}

\appendix
\section{Models for single time series}
We considere the case $N=1$, i.e. when $\mathbf{X}$ is a $1\times T$ matrix or equivalently a $T$-dimensional row vector. Let us denote the entries of $\mathbf{X}$ as $x(t)$.

\subsection{Uniform random walk model}
The trivial model is obtained when no constraints are enforced. In this case, there is no free parameter and the Hamiltonian has the form
\begin{equation}
H(\mathbf{X})=0 
\end{equation} 
As a result, the partition function is 
\begin{equation}
 Z=\sum_\mathbf{X} 1= 2^T
\end{equation} 
which is nothing but the number of possible binary time series of length $T$.
The probability of occurrence  of a time series $\mathbf{X}$ is then
\begin{equation}
P\left(\mathbf{X}\right)=\frac{1}{Z } =2^{-T}
\label{eq:appPrw}
\end{equation}
and is completely uniform over the ensemble of all binary time series of length $T$. 
All the $T$ elements of $\mathbf{X}$ are mutually independent and identically distributed with probability
\begin{equation}
P_t (x)\equiv\textrm{Prob}\big(x(t)=x\big)=\left\{\begin{array}{lr}
{1}/{2}&\quad x=-1\\
{1}/{2}&\quad x=+1\end{array}\right.
\end{equation}
This results in a completely uniform random walk with zero expected value for each increment:
\begin{equation}
\langle x(t)\rangle=0 
\end{equation}
While the (ensemble) variance of each increment equals
\begin{equation}
\textrm{Var}[x(t)]\equiv\langle x^2(t)\rangle-{\langle x(t)\rangle}^2=1.
\end{equation}

\subsection{Biased random walk model}
We now consider the total increment as the simplest non-trivial (one-dimensional)  constraint:
\begin{equation}
C(\mathbf{X})=T\cdot M_1(\mathbf{X})=T\cdot\overline{x(t)}
\label{eq:Cbiasedapp}
\end{equation}
If we denote the corresponding (scalar) Lagrange multiplier by ${\theta}$, the Hamiltonian has the form
\begin{equation}
H\left(\mathbf{X},\theta\right)=\theta \cdot T\cdot \overline{x(t)}
=\theta \sum_{t=1}^T x(t).
\end{equation}
The partition function is 
\begin{eqnarray}  
Z(\theta)&=&\sum_{\mathbf{X}} e^{-\theta\sum_{t=1}^T x(t)} =\sum_{\mathbf{X}} \prod_{t=1}^T e^{-\theta x(t)}\nonumber\\
&=& \prod_{t=1}^T\sum_{x=\pm 1} e^{-\theta x}=\prod_{t=1}^T \left[e^{-\theta} +e^{+\theta}\right]\nonumber\\
&=& \left[ e^{-\theta} +e^{+\theta} \right]^T
\end{eqnarray}
where, when interchanging the order of the sum and product, we have replaced the sum over all time series $\mathbf{X}$ with the sum over the two possible values $x=\pm 1$ of each individual entry.

The probability of the occurrence of a time series $\mathbf{X}$ is
\begin{eqnarray}
 P(\mathbf{X}|\theta)&=& \frac{e^{-\theta \sum_{t=1}^T x(t)}}{\left[ e^{-\theta} +e^{+\theta} \right]^T}=  \prod_{t=1}^T\frac{ e^{-\theta x(t)}}{ e^{-\theta} +e^{+\theta}}\nonumber\\
&=&  \prod_{t=1}^T P_t\big(x(t)|\theta\big)
\end{eqnarray}
where we have introduced the probability $P_t(x|\theta)$  of a given increment $x=\pm 1$ at time $t$, which we identify as
\begin{equation}
P_t(x|\theta)= \frac{ e^{-\theta x}}{ e^{-\theta} +e^{+\theta}}.
\end{equation}
The above expression shows that the stochastic process corresponding to this model is a biased random walk, as the two outcomes $x=\pm 1$ have a different probability, unless $\theta=0$ (which leads us back to the uniform random walk model considered above).

The expected value of the $t$-th increment $x(t)$ (representing the bias of the random walk) is 
\begin{equation}
 \langle x(t)\rangle_\theta=  \sum_{x=\pm 1} x P_t (x|\theta)
=\frac { e^{-\theta}-e^{+\theta}}{e^{-\theta}+e^{+\theta} }=-\tanh{\theta}
\end{equation}
and the variance is 
\begin{equation} 
\textrm{Var}[x(t)] = \langle x^2(t)\rangle_\theta- {\langle x(t)\rangle_\theta}^2=   1- \tanh^2{\theta}.
\end{equation}

The maximum likelihood condition \eqref{eq:match}, fixing the value $\theta^*$ of the parameter $\theta$ given a real time series $\mathbf{X}^*$, reads 
\begin{equation} 
T\big\langle  \overline{x(t)}\big\rangle  =\sum_{t=1}^T\langle x(t)\rangle=-T\tanh{\theta}=T  \cdot\overline{x^*(t)} 
\end{equation}
Where $\overline{x^*(t)}$ is the measured average increment in the observed time series $\mathbf{X}^*$. 
This yields
\begin{equation} 
-\tanh{\theta^*}=\overline{x^*(t)}
\end{equation}
which gives a parameter value
\begin{equation} 
\theta^*=-\textrm{artanh}\left[\overline{x^*(t)}\right]=-\frac{1}{2}\ln\left[\frac{1+\overline{x^*(t)}}{1-\overline{x^*(t)}}\right]
\end{equation}

\subsection{One-dimensional Ising model}
We now consider a model where, besides the constraint on the total increment specified in eq.(\ref{eq:Cbiased}), we enforce an additional constraint on the time-delayed (lagged) quantity $T\cdot B_1(\mathbf{X})$, where $B_1(\mathbf{X})$ is defined in eq.\eqref{eq:B} with $\tau=1$.
This amounts to enforce the average one-step temporal autocorrelation of the time series.
The resulting 2-dimensional constraint can be written as the column vector
\begin{equation}
\vec{C}(\mathbf{X})=\left(\begin{array}{c}C_1(\mathbf{X})\\C_2(\mathbf{X})\end{array}\right)=
T\cdot\left(\begin{array}{c} M_1(\textbf{X})\\ B_1(\textbf{X})\end{array}\right).
\end{equation}
If we write the corresponding Lagrange multiplier as
\begin{equation}
\vec{\theta}=\left(\begin{array}{c}\theta_1\\\theta_2\end{array}\right)=-
\left(\begin{array}{c} I\\K\end{array}\right),
\end{equation}
then the Hamiltonian reads
\begin{eqnarray} 
H(\textbf{X},I,K) &=&\vec{\theta}\cdot\vec{C}(\mathbf{X})
=T\theta_1M_1(\mathbf{X})
+T\theta_2B_1(\mathbf{X})\nonumber\\
&=&-I \sum_{t=1}^T x(t)
 -K \sum^{T}_{t=1} x(t)x(t\!+\!1), \end{eqnarray}
where we consider a periodicity condition as in eq.\eqref{eq:mod} with $\tau=1$, i.e. $x(T+1)\equiv x(1)$. Note that, when $\mathbf{X}$ is a real binary time series of length $T$, this condition can be always enforced by adding one last (fictious) timestep $T+1$ and a corresponding increment $x(T+1)$ chosen equal to $x(1)$. 
For long time series, this has a negligible effect.

The above Hamiltonian coincides with that for the one-dimensional Ising model with periodic boundary conditions \cite{baxter}.
Each time step $t$ is seen as a site in an ordered chain of length $T$, and each value $x(t)=\pm 1$ is seen as the value of a spin sitting at that site.
The model is analytically solvable, which allows us to apply it to real time series in our formalism. 
For the readers familar with time series analysis but not necessarily with the Ising model, we briefly recall the standard solution of the model, adapting it from ref. \cite{baxter}. 

Applying the periodicity condition of eq.\eqref{eq:mod} ensures that all sites (time steps) are statistically equivalent, i.e.:
\begin{equation} 
 \langle x(1)\rangle=\langle x(2)\rangle=\dots=\langle x(T)\rangle
\end{equation}
so that the system is translationally (here, temporally) invariant. 
The partition function is  
\begin{equation} 
Z(I,K) =  \sum_{\mathbf{X}} \exp\left[I \sum^{T}_{t=1} x(t) + K \sum^{T}_{t=1}x(t)x(t\!+\! 1) \right]\nonumber
\end{equation}
and can be rewritten as a product of terms involving only two successive time steps:
\begin{equation} 
Z(I,K) =\sum_{\mathbf{X}} \prod_{t=1}^T V\big(x(t),x(t+1)\big),
\label{eq:partition}
\end{equation} 
where we have introduced the function $V(x,y)$ defined as
\begin{equation} 
V(x,y)\equiv\exp\left(I\frac{x+y}{2}+Kxy\right).
\end{equation}

We since both $x$ and $y$ can take only the values $\pm 1$, we can regard $V(x,y)$ as the element of a $2\times2$ matrix $\mathbf{V}$ called the \emph{transfer matrix} \cite{baxter}:
\begin{equation} 
\mathbf{V}\equiv \left( \begin{array}{cc} V(+1,+1) &  V(+1,-1) \\ V(-1,+1)&  V(-1,-1) \end{array}\right) =  \left( \begin{array}{cc} e^{K+I}&  e^{-K} \\  e^{-K}&  e^{K-I} \end{array}\right).
\end{equation} 
This allows us to rewrite eq.\eqref{eq:partition} as 
\begin{equation} 
Z(I,K) = \textrm{Tr} \big(\mathbf{V}^T\big).
\label{eq:trace}
\end{equation}
Let $\vec{v}_1, \vec{v}_2$ denote the two eigenvectors of $\mathbf{V}$, and $\lambda_1 ,\lambda_2$ the corresponding eigenvalues, so that 
\begin{equation}
\mathbf{V}\vec{v}_j=\lambda_j\vec{v}_j,\quad j=1,2.
\end{equation}
The $2\times 2$ matrix $\mathbf{Q}\equiv (\vec{v}_1,\vec{v}_2)$ (having column vectors $\vec{v}_1$ and $\vec{v}_2$) diagonalizes $\mathbf{V}$, i.e.  
\begin{equation} 
\mathbf{V}=\mathbf{Q}  \left( \begin{array}{cc} \lambda_1 &  0 \\ 0&  \lambda_2 \end{array}\right) \mathbf{Q}^{-1},
\end{equation}
where a direct calculation of the eigenvalues and eigenvectors yields
\begin{eqnarray}
\lambda_1 &=& e^K \cosh I+\sqrt{e^{2K}\sinh^2 I+e^{-2K}}\\
\lambda_2 &=& e^K \cosh I-\sqrt{e^{2K}\sinh^2 I+e^{-2K}}
\end{eqnarray}
and
\begin{equation} 
 \mathbf{Q}=\left( \begin{array}{cc} \cos{\phi} & -\sin{\phi} \\
\sin{\phi}& \cos{\phi} \end{array} \right),
\end{equation}
with $\phi$ defined by
\begin{equation} 
\cot {2\phi} \equiv e^{2K}\sinh I.
\label{eq:phi}
\end{equation}
It then follows that eq.\eqref{eq:trace} simply reduces to 
\begin{equation} 
Z(I,K) = \textrm{Tr}\left( \begin{array}{cc} \lambda_1 &  0 \\ 0&  \lambda_2 \end{array}\right)^{T} = \lambda_1^{T} +\lambda_2^{T},
\end{equation}
and the probability of occurrence of a time series $\mathbf{X}$ is 
\begin{equation}
P(\mathbf{X}|I,K)=\frac{\prod_{t=1}^T V\big(x(t),x(t+1)\big)}{\lambda_1^{T} +\lambda_2^{T}}.
\label{eq:twostep}
\end{equation}

The above results allow us to analytically obtain expected values. That of $x(t)$ is
\begin{eqnarray} 
\langle x(t)\rangle=
\sum_\mathbf{X}x(t)P(\mathbf{X}|I,K)=
 \frac{\textrm{Tr}\big(\mathbf{S}\mathbf{V}^T\big)}{\lambda_1^{T} +\lambda_2^{T}},
\label{eq:x}
\end{eqnarray}
where we have introduced the diagonal matrix
\begin{equation}
 \mathbf{S}\equiv\left( \begin{array}{cc} S(+1,+1) &  S(+1,-1) \\ S(-1,+1)&  S(-1,-1) \end{array}\right) 
=\left( \begin{array}{rr} +1 & 0 \\
0& -1 \end{array} \right)
\end{equation}
having elements 
\begin{equation}
S(x,y)\equiv x \delta (x,y).
\end{equation}
Similarly, for $0<s-t<T$ the expected value of $x(t)x(s)$ is  
\begin{eqnarray} 
\langle x(t)x(s)\rangle&=&
\sum_\mathbf{X}x(t)x(s)P(\mathbf{X}|I,K)\nonumber\\
&=& \frac{\textrm{Tr} \big( \mathbf{S}\mathbf{V}^{s-t}\mathbf{S}\mathbf{V}^{T+t-s}\big)}{\lambda_1^{T} +\lambda_2^{T}}.
\label{eq:xx}
\end{eqnarray}
In the limit $T\to\infty$ (corresponding to long time series in our case) with $s-t$ fixed, these expressions become 
\begin{eqnarray}
\langle x(t)\rangle&=& \cos{2\phi} \\
\langle x(t) x(s) \rangle&=& \cos^2{2\phi}+ \sin^2{2\phi} \left( \frac{\lambda_1}{\lambda_2}\right)^{s-t}
\end{eqnarray}
Now, we note that eqs.\eqref{eq:x} and \eqref{eq:xx} manifestly show the translational (temporal) invariance of the model, as $\langle x(t)\rangle$ is independent of $t$ and   $\langle x(t)x(s)\rangle$ depends on $t$ and $s$ only through their difference $s-t$. 
This implies that, writing $\tau\equiv s-t$ and performing a temporal average,
\begin{eqnarray}
\big\langle M_1\big\rangle&=&\cos{2\phi} \\
\big\langle B_\tau\big\rangle&=& \cos^2{2\phi}+ \sin^2{2\phi} \left( \frac{\lambda_1}{\lambda_2}\right)^{\tau}.
\end{eqnarray}
Using eq.\eqref{eq:phi} we can rewrite these expressions in terms of the model parameters, $I$ and $K$, as
\begin{eqnarray}
\big\langle M_1\big\rangle&=&
 \frac{e^{2K}\sinh{I}}{\sqrt{1+e^{4K}\sinh^2{I}}} \\
\big\langle B_\tau\big\rangle&=&
 \frac{e^{4K}\sinh^2{I}+(\lambda_1/\lambda_2)^\tau}{1+e^{4K}\sinh^2{I}}.
\end{eqnarray}

The expected value of the autocorrelation defined in eq. \eqref{eq:autocorr} can be approximated as the ratio of two expected values as follows:
\begin{equation}
\big\langle A_\tau\big\rangle\equiv
\left\langle\frac{B_\tau- M_1^2}{1- M_1^2}\right\rangle\approx\frac{\langle B_\tau\rangle-\langle M_1^2\rangle}{1-\langle M_1^2\rangle}= \left( \frac{\lambda_1}{\lambda_2}\right)^\tau.
\end{equation}

\section{Models for single cross-sections of multiple time series}
For a single cross-section of a set of $N$ multiple time series, $\mathbf{X}$ is a $N\times 1$ matrix or equivalently a $N$-dimensional column vector. 
We denote the entries of $\mathbf{X}$ as $x_i$.

\subsection{Uniform random walk model}
The uniform random walk is a simple modification of the same model that we considered for single time series, where $x(t)$ is replaced by $x_i$ and $T$ is replaced by $N$.
This model is obtained when no constraints are enforced. The Hamiltonian is
\begin{equation}
H(\mathbf{X})=0 
\end{equation} 
and the partition function is simply the number of possible configurations for a single cross-section of $N$ stocks:
\begin{equation}
 Z=\sum_\mathbf{X} 1= 2^N.
\end{equation} 
The probability of occurrence  of a cross section $\mathbf{X}$ is
\begin{equation}
P\left(\mathbf{X}\right)=\frac{1}{Z } =2^{-N}
\label{eq:appPcross}
\end{equation}
and is completely uniform over the ensemble of all cross sections of $N$ stocks. 
All the $N$ elements of $\mathbf{X}$ are mutually independent and identically distributed with probability
\begin{equation}
P_i (x)\equiv\textrm{Prob}\big(x_i=x\big)=\left\{\begin{array}{lr}
{1}/{2}&\quad x=-1\\
{1}/{2}&\quad x=+1\end{array}\right.
\end{equation}
This results in a completely uniform random walk with zero expected value
\begin{equation}
\langle x_i\rangle=0 
\end{equation}
and maximum variance
\begin{equation}
\textrm{Var}[x_i]\equiv\langle x_i^2\rangle-{\langle x_i\rangle}^2=1.
\end{equation}

\subsection{Biased random walk model}
Also this model is analogous to the corresponding model for single time series.
We select the total daily increment of the cross section $\mathbf{X}$ as the constraint:
\begin{equation}
C(\mathbf{X})=N\cdot M_1(\mathbf{X})=N\cdot\{x_i\}
\end{equation}
Let the corresponding Lagrange multiplier be denoted by ${\theta}$. The Hamiltonian is
\begin{equation}
H\left(\mathbf{X},\theta\right)=\theta \cdot N\cdot \{x_i\}
=\theta \sum_{i=1}^N x_i
\end{equation}
and the partition function is 
\begin{eqnarray}  
Z(\theta)&=&\sum_{\mathbf{X}} e^{-\theta\sum_{i=1}^N x_i} =\sum_{\mathbf{X}} \prod_{i=1}^N e^{-\theta x_i}\nonumber\\
&=& \prod_{i=1}^N\sum_{x=\pm 1} e^{-\theta x}=\prod_{i=1}^N \left[e^{-\theta} +e^{+\theta}\right]\nonumber\\
&=& \left[ e^{-\theta} +e^{+\theta} \right]^N.
\end{eqnarray}

The probability of the occurrence of a cross section $\mathbf{X}$ is
\begin{eqnarray}
 P(\mathbf{X}|\theta)&=& \frac{e^{-\theta \sum_{i=1}^N x_i}}{\left[ e^{-\theta} +e^{+\theta} \right]^N}=  \prod_{i=1}^N\frac{ e^{-\theta x_i}}{ e^{-\theta} +e^{+\theta}}\nonumber\\
&=&  \prod_{i=1}^N P_i\big(x_i|\theta\big)
\label{eq:appPcrossbias}
\end{eqnarray}
where we have introduced the probability $P_i(x|\theta)$  of a given increment $x=\pm 1$ for stock $i$, which we identify as
\begin{equation}
P_i(x|\theta)= \frac{ e^{-\theta x}}{ e^{-\theta} +e^{+\theta}}.
\end{equation}
Just like the corresponding model for single time series, this model is a biased random walk, because the two outcomes $x=\pm 1$ have a different probability unless $\theta=0$.

The expected value of the $i$-th increment $x_i$ is 
\begin{equation}
 \langle x_i\rangle_\theta=  \sum_{x=\pm 1} x P_i (x|\theta)
=\frac { e^{-\theta}-e^{+\theta}}{e^{-\theta}+e^{+\theta} }=-\tanh{\theta}
\end{equation}
and the variance is 
\begin{equation} 
\textrm{Var}[x_i] = \langle x^2_i\rangle_\theta- {\langle x_i\rangle_\theta}^2=   1- \tanh^2{\theta}.
\end{equation}

The maximum likelihood condition \eqref{eq:match}, fixing the value $\theta^*$ of the parameter $\theta$ given a real cross section $\mathbf{X}^*$, reads 
\begin{equation} 
N\big\langle  \{x_i\}\big\rangle  =\sum_{i=1}^N\langle x_i\rangle=-N\tanh{\theta}=N  \cdot\{x^*_i\} 
\end{equation}
where $\{x^*_i\}$ is the measured average increment of the observed cross section $\mathbf{X}^*$. 
This yields
\begin{equation} 
-\tanh{\theta^*}=\{x^*_i\}
\end{equation}
which gives a parameter value
\begin{equation} 
\theta^*=-\textrm{artanh}\left[\{x^*_i\}\right]=-\frac{1}{2}\ln\left[\frac{1+\{x^*_i\}}{1-\{x^*_i\}}\right].
\end{equation}

\subsection{Mean-field Ising model}
In this model, we enforce two constraints: the total increment and the total coupling between stocks.
The resulting 2-dimensional constraint can be written as
\begin{equation}
\vec{C}(\mathbf{X})=\left(\begin{array}{c}C_1(\mathbf{X})\\C_2(\mathbf{X})\end{array}\right)=
\left(\begin{array}{c} N\cdot M_1(\textbf{X})\\ D(\textbf{X})\end{array}\right).
\end{equation}
We can write the corresponding Lagrange multiplier as
\begin{equation}
\vec{\theta}=\left(\begin{array}{c} \theta_1\\ \theta_2\end{array}\right)
=-\left(\begin{array}{c} h\\ J\end{array}\right)
\end{equation}
and the Hamiltonian as
\begin{equation}
H(\mathbf{X},h,J)= -h\sum_{i=1}^{N} x_i - J\sum_{i<j}x_i x_j.
\end{equation}

Note that here we are not enforcing nearest-neighbor interactions as in the one-lagged model for single time series, but market-wide interactions among all stocks for the same time step (cross section). This is the result of the fact that, when considering cross sections, there is no natural notion of `lattice sites' induced by e.g. a temporal ordering as in the one-lagged model. In other words, pairs of stocks in a cross section are neither `close' nor `distant'. We therefore assume a common interaction strength $J$ among all stocks.

The above model, known as the mean-field Ising model, is analytically solvable. Here we adapt the derivation illustrated in ref. \cite{baxter}.
We first note that, since ${x_i}^2=1$ for all $i$, $H(\mathbf{X},h,J)$ can be expressed as a function of $M_1(\textbf{X})$ alone:
\begin{equation}
H(\mathbf{X},h,J)=-hNM_1(\textbf{X})-\frac{J}{2}\big[N^2M^2_1(\textbf{X})-N\big].
\label{eq:HM}
\end{equation}
This implies that the sum over configurations in the partition function can
be replaced by a sum over the allowed values of $M_1(\textbf{X})$, weighted by the number of configurations for each value.
If we denote by $r$ the number of increments that are negative $(x=-1)$, and by $(N-r)$ the number of increments that are positive $(x=+1)$, then we can write the Hamiltonian as a function of $r$ alone through the expression
\begin{equation}
N M_1(\textbf{X})=N-2r.
\end{equation}
The partition function can therefore be calculated as
\begin{equation}
Z(h,J) \equiv \sum_{\mathbf{X}} e^{-H(\mathbf{X},h,J)}=\sum_{r=1}^{N} C_r
\end{equation}
where 
\begin{equation}
 C_r\equiv  \frac {N!}{r! (N-r)!} e^{h(N-2r)+\frac{J}{2}[{(N-2r)}^2-N]}
\end{equation}
incorporates the binomial coefficient enumerating the configurations with given $r$.
The expected increment is therefore
\begin{equation}
\langle M_1\rangle=\left\langle1-\frac{2r}{N}\right\rangle=\frac{ \sum_{r=1}^{N}\left(1-\frac{2r}{N}\right) C_r}{Z(h,J)}\quad \forall i.
\label{eq:expM}
\end{equation}

When $N$ is large, a traditional derivation \cite{baxter} shows that the sum at the numerator of eq.\eqref{eq:expM} is dominated by the single addendum corresponding to the maximum of $C_r$. The same applies to the partition function at the denominator. If $r_0$ denotes the value of $r$ such that $C_r$ is maximum, we then get
\begin{equation}
\langle M_1\rangle\approx 1-\frac{2r_0}{N}.
\label{eq:M1bis}
\end{equation}
A further expansion \cite{baxter} finally shows that, given $h$ and $J$, the expected value $\langle M_1\rangle$ is the solution of the nonlinear equation
\begin{equation} 
\langle M_1\rangle=  \tanh\big[(N-1)J\langle M_1\rangle+h\big].
\end{equation}
From the above equation, one can infer the existence of a phase transition in the model, separating a regime where the expected `magnetization' (here the average increment $\langle M_1\rangle$) is zero from one where it is non-zero \cite{baxter}. This transition is discussed in sec.\ref{sec:crossmfm}.

Before proceeding further, we note a peculiarity of the model, which has implications for the applicability of our maximum likelihood approach.
An argument similar to that leading to eq.\eqref{eq:M1bis} implies that the second moment of $M_1(\mathbf{X})$ can be expressed as
\begin{equation}
\langle M^2_1\rangle=\left\langle\left(1-\frac{2r}{N}\right)^2\right\rangle\approx \left(1-\frac{2r_0}{N}\right)^2\approx \langle M_1\rangle^2.
\label{eq:M2}
\end{equation}
This implies that
\begin{equation}
\textrm{Var}[M_1]\equiv \langle M_1^2\rangle-\langle M_1\rangle^2=0,
\end{equation}
or in other words that $M_1(\mathbf{X})$ is no longer a random variable.
As a consequence, something unusual happens when we apply the maximum likelihood principle.
From eq.\eqref{eq:HM}, and recalling the general result embodied by eq.\eqref{eq:match2} in sec.\ref{sec:ML}, it is clear that the parameter values $h^*$ and $J^*$ maximizing the likelihood can be found as the solution to the two coupled equations
\begin{eqnarray}
\langle M_1\rangle&=&M_1(\mathbf{X}^*)\label{eq:mm1}\\
\langle M_1^2\rangle&=&M^2_1(\mathbf{X}^*)\label{eq:mm2}
\end{eqnarray}
However, eq. \eqref{eq:M2} implies that eq.\eqref{eq:mm2} can be rewritten as 
\begin{eqnarray}
\langle M_1\rangle^2=M^2_1(\mathbf{X}^*)
\label{eq:trivial}
\end{eqnarray}
which coincides with eq.\eqref{eq:mm1}. So eqs. \eqref{eq:mm1} and \eqref{eq:mm2} are equivalent, and they cannot be used to uniquely determine the two unknown parameters $h^*$ and $J^*$.
This is the result of the fact that, when fitted to the data, the model is actually over-constrained: there are two parameters to fit the only constraint ($M_1$) on which the Hamiltonian depends.
This aspect of the model is not manifest when $M_1$ is regarded as a function of $h$ and $J$, as usually done when simulating spin systems. 

The above consideration implies that we should drop one of the two parameters and consider the two cases $J=0$ and $h=0$ separately.
The former case coincides with the biased random walk model that we already discussed, and we will not discuss it any further. 
The latter case will instead represent our genuine specification of the `mean-field' model.
Setting $h=0$ implies 
\begin{equation}
H(\mathbf{X},0,J)=-\frac{J}{2}\big[N^2M^2_1(\textbf{X})-N\big]
\label{eq:HM2}
\end{equation}
and
\begin{equation} 
\langle M_1\rangle=  \tanh\big[(N-1)J\langle M_1\rangle\big].
\label{eq:nonlinear}
\end{equation}
Applying the maximum likelihood principle to eq.\eqref{eq:HM2} tells us to select $J^*$ as the solution of eq.\eqref{eq:mm2}. 
However, we have seen that this condition leads to eq.\eqref{eq:trivial}, which is actually equivalent to eq.\eqref{eq:mm1}.
Therefore, the value of $J^*$ can be found by replacing $\langle M_1\rangle$ with the observed value $M_1(\mathbf{X}^*)=\{x^*_i\}$ in eq. \eqref{eq:nonlinear}, which leads to
\begin{equation} 
\{x^*_i\}=  \tanh\big[(N-1)J^*\{x^*_i\}\big].
\label{eq:nonlinear2}
\end{equation}
Note that in the traditional situation one is interested in finding the (expected) magnetization given a value of $J$, which implies that the transcendental eq. \eqref{eq:nonlinear} should be solved numerically.
Here, we are instead facing the inverse situation where we look for the value of $J^*$ given the (observed) value of the magnetization. 
In this quite unusual case, it turns out that eq. \eqref{eq:nonlinear2} can be inverted to give the following analytical solution:
\begin{equation} 
J^*=\frac{\textrm{artanh}\{x^*_i\}}{\{x^*_i\}(N-1)}
=\frac{1}{2\{x^*_i\}(N-1)}\ln\left[\frac{1+\{x^*_i\}}{1-\{x^*_i\}}\right].
\label{eq:fixJ}
\end{equation}
Once this value is calculated, it can be inserted into the probability  
\begin{equation}
P(\mathbf{X}^*|0,J)= \frac{e^{-H(\mathbf{X}^*,0,J)}}{Z(0,J)}= \frac { e^{JN(N\{x^*_i\}^2-1)/2 }}{ \sum_{r=1}^{N}  \frac {N!}{r! (N-r)!} e^{J[{(N-2r)}^2-N]/2} }
\end{equation}
(where we have set $h=0$) to obtain the maximized likelihood of generating the observed cross section $\mathbf{X}^*$ under the mean-field model.

%%%
%\input acknowledgement.tex   % input acknowledgement

\end{document}